\documentclass[aps,prb,twocolumn,showpacs,superscriptaddress,longbibliography,floatfix,nofootinbib]{revtex4-2}

\usepackage{amsmath, amsfonts, amssymb, bm}
\usepackage{mathtools}
\usepackage[pdftex]{graphicx}
\usepackage{import}
\usepackage{ulem}

\usepackage[breaklinks,hypertexnames=false]{hyperref}
\hypersetup{colorlinks,linkcolor={magenta},citecolor={blue},urlcolor={blue}} 

\usepackage{braket}
\usepackage{lipsum}
\usepackage{xcolor}

\usepackage[capitalize]{cleveref}

\usepackage{glossaries}
\glsdisablehyper

\newacronym{mbs}{MBS}{Majorana bound state}
\newacronym{mzm}{MZM}{Majorana zero mode}
\newacronym{abs}{ABS}{Andreev bound state}
\newacronym{fi}{FI}{ferromagnetic insulator}
\newacronym{jj}{JJ}{Josephson junction}
\newacronym{cpr}{CPR}{current-phase relation}

\usepackage[margin=0.9in]{geometry}
\usepackage[utf8]{inputenc}
\usepackage[inline,shortlabels]{enumitem}
\usepackage{longtable}
\usepackage[caption=false]{subfig}
\usepackage{multirow}

\usepackage{sidecap,tikz}
\definecolor{lime}{HTML}{A6CE39}
\DeclareRobustCommand{\orcidicon}{\hspace{-1mm}
	\begin{tikzpicture}
		\draw[lime, fill=lime] (0,0) 
		circle [radius=0.16] 
		node[white] {{\fontfamily{qag}\selectfont \tiny \,ID}};
		\draw[white, fill=white] (-0.0525,0.095) 
		circle [radius=0.007];
	\end{tikzpicture}
	\hspace{-3mm}
}
\foreach \x in {A, ..., Z}{\expandafter\xdef\csname orcid\x\endcsname{\noexpand\href{https://orcid.org/\csname orcidauthor\x\endcsname}
		{\noexpand\orcidicon}}
}

\def\be{\begin{equation}}
\def\ee{\end{equation}}
\def\bea{\begin{eqnarray}}
\def\eea{\end{eqnarray}}
\def\bmat{\begin{pmatrix}}
\def\emat{\end{pmatrix}}
\def\bs{\begin{split}}
\def\es{\end{split}}

\def\~{$\approx$}

\def\dag{\dagger}

\newcommand{\up}{\uparrow}
\newcommand{\dw}{\downarrow}

\let\Re\undefined
\let\Im\undefined
\DeclareMathOperator{\Re}{\mathfrak{Re}}
\DeclareMathOperator{\Im}{\mathfrak{Im}}
\DeclareMathOperator{\diag}{diag}

\begin{document}

\title{
Interplay between Majorana and Shiba states in a minimal Kitaev chain coupled to a superconductor
}

\author{M. Alvarado\orcidA{}}
\affiliation{ 
Instituto de Ciencia de Materiales de Madrid (ICMM), Consejo Superior de Investigaciones Cient{\'i}ficas (CSIC), Sor Juana In{\'e}s de la Cruz 3, 28049 Madrid, Spain}
\affiliation{ 
Departamento de F{\'i}sica Te{\'o}rica de la Materia Condensada, Condensed Matter Physics Center (IFIMAC)
and Instituto Nicol{\'a}s Cabrera, Universidad Aut{\'o}noma de Madrid, 28049 Madrid, Spain}

\author{A. Levy Yeyati\orcidD{}}
\affiliation{ 
Departamento de F{\'i}sica Te{\'o}rica de la Materia Condensada, Condensed Matter Physics Center (IFIMAC)
and Instituto Nicol{\'a}s Cabrera, Universidad Aut{\'o}noma de Madrid, 28049 Madrid, Spain}

\author{Ram\'{o}n Aguado\orcidC{}}
\affiliation{ 
Instituto de Ciencia de Materiales de Madrid (ICMM), Consejo Superior de Investigaciones Cient{\'i}ficas (CSIC), Sor Juana In{\'e}s de la Cruz 3, 28049 Madrid, Spain}

\author{R. Seoane Souto\orcidB{}}
\affiliation{ 
Instituto de Ciencia de Materiales de Madrid (ICMM), Consejo Superior de Investigaciones Cient{\'i}ficas (CSIC), Sor Juana In{\'e}s de la Cruz 3, 28049 Madrid, Spain}

\date{\today}

\begin{abstract}
Two semiconducting quantum dots (QDs) coupled through a superconductor constitute a minimal realisation of a Kitaev chain with Majorana zero modes (MZMs). Such MZMs can be detected by {\it e.g.}, tunneling conductance between each QD and normal leads [\href{https://doi.org/10.1038/s41586-022-05585-1}{Dvir et al, Nature {\bf 614}, 445 (2023)}]. We here discuss how the seemingly trivial substitution of one of the normal leads by a superconducting (SC) one gives rise to a plethora of new effects. In particular, the coupling to the SC lead induces non-local Majorana effects upon variations of the QDs' energies. Furthermore, the lowest excitation of the chain is no longer determined by the bulk gap but rather by the energy of an emergent subgap Yu-Shiba-Rusinov (YSR) state coexisting with the MZMs. The YSR state hybridizes with the MZMs when the coupling between the SC and the QD is larger than the spin splitting, spoiling the Majorana properties, including the quantized conductance.
\end{abstract}

\maketitle

\section{Introduction}\label{SecI}
The Kitaev chain~\cite{Kitaev2001} is the minimal model for one-dimensional spinless p-wave superconductivity. In the topological phase, the chain hosts  Majorana Zero Modes (MZMs)~\cite{Wilczek2009, Alicea2012, Leijnse2012_b, Aguado2017, Beenakker2020, Tanaka2024} at its ends, which feature non-abelian statistics that can be exploited for topological quantum information processing~\cite{Nayak2008, Sarma2015, Lahtinen2017,Aguado2020}. New platforms based on arrays of quantum dots (QDs) connected through mesoscopic superconductors~\cite{Leijnse2012, Sau2012, Fulga2013, Leijnse2024} have emerged as a promising route toward bottom-up engineering of artificial Kitaev chains. This approach largely avoids undesired effects associated to disorder and material inhomogeneities inherent to other platforms~\cite{Prada2012, Cayao2015, Liu2017, Prada2020}.

A minimal Kitaev chain can be realized by just two QDs coupled through a mesoscopic superconductor that mediates crossed Andreev reflection (CAR) and elastic cotunneling (ECT) between the QDs~\cite{Leijnse2012, Sau2012}. Such minimal Kitaev chains host ``poor man's Majorana states'', that share properties with their topological counterparts~\cite{Tsintzis2024}, but lack topological protection, since they appear at fine-tuned ``sweet spots'' in parameter space with equal CAR and ECT. Recent experiments have demonstrated exquisite control on CAR and ECT amplitudes~\cite{Wang2022, Wang2023, Bordin2022, Bordin2023}, which allows to tune sweet spots~\cite{Liu2022, Dvir2023, Bordin2023, Haaf2023, Bordin2024} where one MZM localizes in each QD of the minimal chain. Interestingly, longer chains, already at the three QD level, start to show a bulk gap and some degree of protection~\cite{Bordin2024, Haaf2024}.

\begin{figure*}[t!]
\centering
\includegraphics[width=0.8\textwidth]{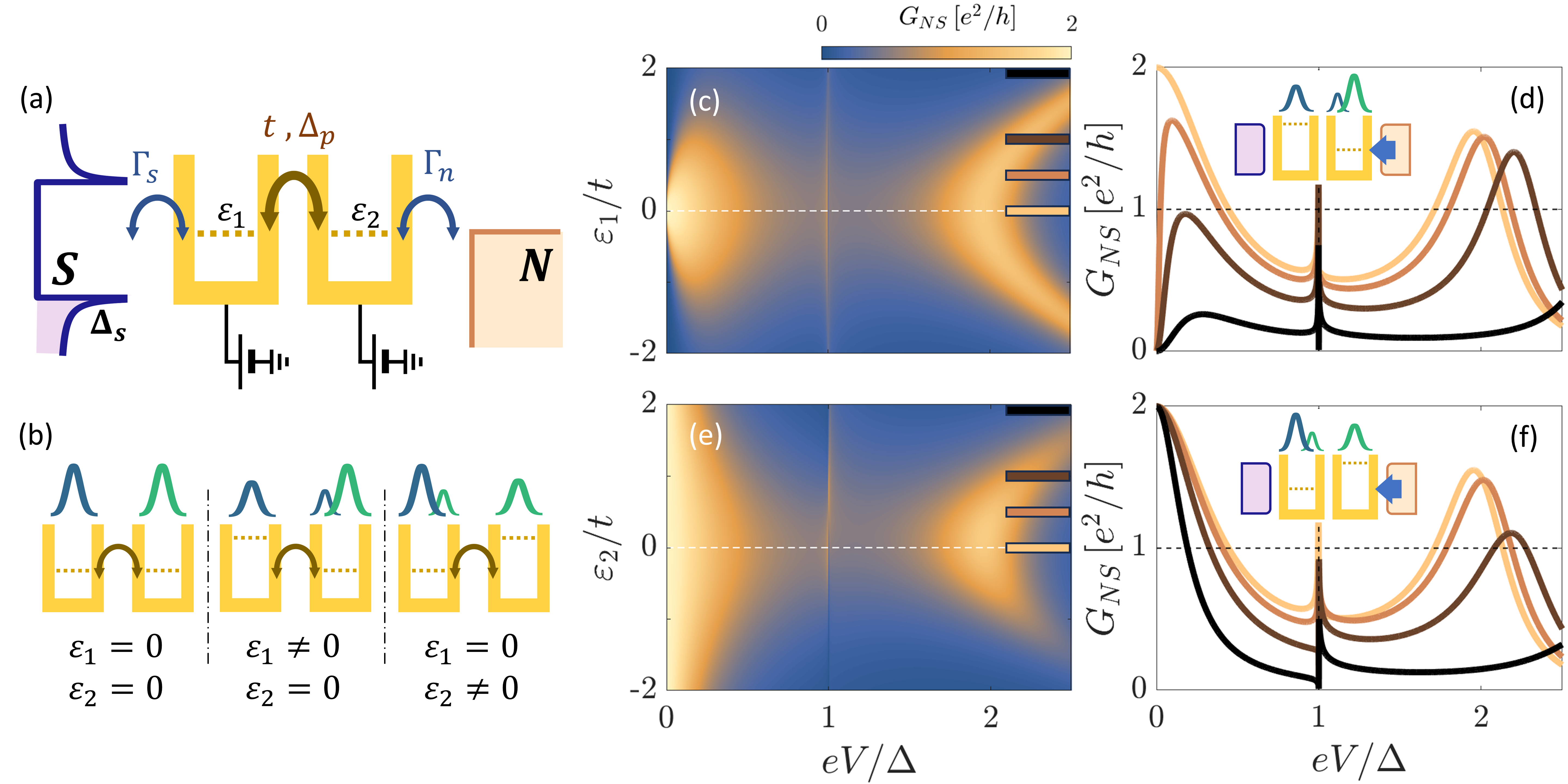}
\caption{(a) Sketch of a minimal Kitaev chain contacted to a normal (N) and a superconducting (S) lead, with tunneling strengths $\lambda_n$ and $\lambda_s$ respectively. (b) MZMs localization in the two QDs (blue and green) for different parameter regimes, assuming $t=\Delta_p$. (c-f) Normal conductance as a function of the QDs levels. Upper (Lower) panel shows the normal conductance as a function of $\varepsilon_1$ ($\varepsilon_2$). (d, f) Conductance cuts for different values of the quantum dots levels $\varepsilon/t= 0,0.5,1,2$. Inset shows a diagram of the setup, where the blue arrow marks the direction of the probing. We set $\Gamma_n/t=0.4$ and $\Gamma_s/t = 0.1$ in all panels. 
}
\label{fig1}
\end{figure*}

So far, the assessment of sweet spots has been done via local and non-local normal conductance, by coupling the system to metallic leads~\cite{Wang2022, Wang2023, Bordin2022, Liu2022, Dvir2023, Bordin2023,Haaf2023,Bordin2024}. In this work, we explore what happens beyond this minimal setup and theoretically study the transport properties of a minimal Kitaev chain coupled to a normal and a superconducting (SC) probe, see Fig.\ref{fig1}(a), inspired by recent experimental configurations, Refs.~\cite{Bordin2023, Bordin2024}. 

One consequence of the replacement of one of the normal leads by a SC one, is the change of the well-known quantized zero-bias peak $G_0 = 2e^2/h$~\cite{Law2009, Flensberg2010, Wimmer2011},  which now becomes $G_s(e|V|=\Delta)=(4-\pi)G_0$ when the BCS coherence peak aligns with the MZM~\cite{Peng2015, Zazunov2016}. Despite the lack of topological protection in this minimal setup, we find a less obvious result: the universal conductance values through either the normal or the SC probe are robust against local detuning of the QD in contact with such probe. 
Further analysis in this work includes a detailed study of the role of the asymmetries and non-local effects induced by the inclusion of the non-trivial self-energy associated to the SC lead. As we discuss, this phenomenology could be used as a way to measure the degree of localization~\cite{Seoane2023} of the MZMs in the minimal Kitaev chain and to characterize sweet spots. 

Interestingly, the regime of strong hybridization between the SC and the minimal Kitaev chain is governed by the emergence of Yu-Shiba-Rusinov (YSR) states~\cite{Yu1965, Shiba1968, Rusinov1969, Sakurai1970, Byers1997, Aguado2015}. Qubits based on subgap and YSR states \cite{Simon2021, Hays2021, Bargerbos2022, Pita2023, Bargerbos2023b, Pita2024, Cheung2024} have been shown to feature enhance robustness against noise \cite{Ruben2024, Gorm2024, Zatelli2024}. Additionally, YSR states have been found to increase the stability of minimal Kitaev chains using toy models that neglect the SC continuum of states~\cite{Miles2023, Liu2023}. Their inclusion, however, results in important physical consequences, since the energy of the YSR states decreases when increasing the coupling to the lead, thus reducing the effective gap between MZMs and the excited states. A finite spin-polarization in the dots generates a destructive interference between the YSR states and the MZMs, which spoils the topological properties of the latter, including the conductance quantization. In the language of non-Hermitian open quantum systems, we understand the topological properties of the hybrid junction \cite{Lado2024, Beenakker2024, Ohnmacht2024} in terms of Exceptional Point (EP) bifurcations in the subgap spectrum~\cite{Berry2004, SanJose2016, Avila2019, Nagai2020, Cayao2024, Pino2024, Cayao2024b}, characterized by a change of the MZMs localization.

The rest of the manuscript is organized as follows: in Sec.~\ref{SecII} we introduce the effective minimal model to describe the system, giving some analytical insights on the renormalization of the excited state by means of the coupling to the SC lead. 

In Sec.~\ref{SecIII} and \ref{SecIV} we study both, the normal and superconducting conductance when detuning both QDs in the weak and strong coupling regime with the SC lead respectively. Sec.~\ref{SecV} is devoted to the analysis of the coexistence of MZMs and YSR states, referring to spectral properties such the localization of the states and the pole structure of the system. We extend this analysis to the finite polarization case in Sec.~\ref{SecVI}, where the destructive interference between MZMs and YSR states spoils the Majorana properties for finite values of the coupling to the SC lead.

We finally offer, in Sec.~\ref{SecVII}, some conclusions, summarizing the main results. Technical details like the explicit analysis of the poles of the system including the exceptional points, some notes on the Green's functions used to compute both, spectral and transport properties and, the extension of the study of the physics associated to the QDs chemical potential to large couplings, are included in the appendices. 

\section{Modelization}\label{SecII}
We examine a minimal Kitaev chain coupled to a normal and a SC leads, as sketched in Fig.\ref{fig1}(a). We consider that the system is subject to a strong magnetic field that fully polarizes the quantum dots spins. The two QDs are coupled via a central SC, that mediates ECT and CAR, with effective amplitudes $t$ and $\Delta_p$ respectively, realizing a one dimensional spinless p-wave SC that can host MZMs~\cite{Leijnse2012}. Written 
in the $4\times4$ (hat notation) Nambu basis $\hat{\Psi} = (\psi_\up, \psi_\up^\dagger, \psi_\dw, \psi_\dw^\dagger)^T$,  the toy model for the minimal Kitaev chain takes the form
\be 
\hat{H}_{\nu} = \varepsilon_{\nu} \sigma_z \tau_0 + 2E_{z,\nu
} \sigma_z \Pi_{\dw}  \ , \quad \hat{T} = (t\sigma_z + i\Delta_p \sigma_y)\Pi_{\up} \ ,
\ee 
where $\nu={1,2}$ corresponds to the different sites in the chain, $\tau_\mu$ ($\sigma_\mu$) are Pauli matrices acting in spin (electron/hole) space, and $\Pi_{\up, \dw} = (\tau_0\pm\tau_z)/2$ are the projectors over the spin up/down sector. Notice that as one reaches the polarized limit of the minimal Kitaev chain when the Zeeman splitting $E_{z,\nu}\rightarrow\infty$, the coupling between the QDs in the spin down sector goes correspondingly to zero, cf. App.~\ref{App_A}. External gates can be used to tune the chemical potential on the two QDs independently ($\varepsilon_1$ and $\varepsilon_2$), changing the degree of localization of the MZMs, see Fig.\ref{fig1}(b). The minimal Kitaev chain is coupled to the SC (normal) lead through the tunneling rate $\Gamma_s = \lambda_s^2/t_s$ ($\Gamma_n = \lambda_n^2/t_n$), being $\lambda_s$ ($\lambda_n$) the coupling and $t_s$ ($t_n$) the bandwidth. 

To describe the transport properties of the device, we use the non-equilibrium Green's function formalism~\cite{Cuevas1996, Zazunov2016, Alvarado2020, Alvarado2021, Alvarado2022}, that allows us to consider arbitrary coupling strengths between the minimal Kitaev chain and the leads, see Appendices~\ref{App_A}, \ref{App_D} and \ref{App_F} for details. In addition, the poles of the advanced Green's function determine the energy of the system's states that can be related to the main transport features. Notice that for a spin-polarized chain, there are no subgap Andreev reflections with the SC probe \cite{Cuevas2020} due to the incompatibility between order parameters across the junction~\cite{Zazunov2012, Peng2015, Zazunov2016} ({\it i.e.}, a BCS lead with s-wave pairing $\Delta_s$ and a spinless SC with p-wave pairing $\Delta_p$~\cite{footnote1, Cayao2024c}). Unless otherwise stated, we use $\Delta_p=\Delta_s=\Delta=t=1$ and $t_s=t_n=10 \, t$.

\begin{figure}[t!]
\centering
\includegraphics[width=\columnwidth]{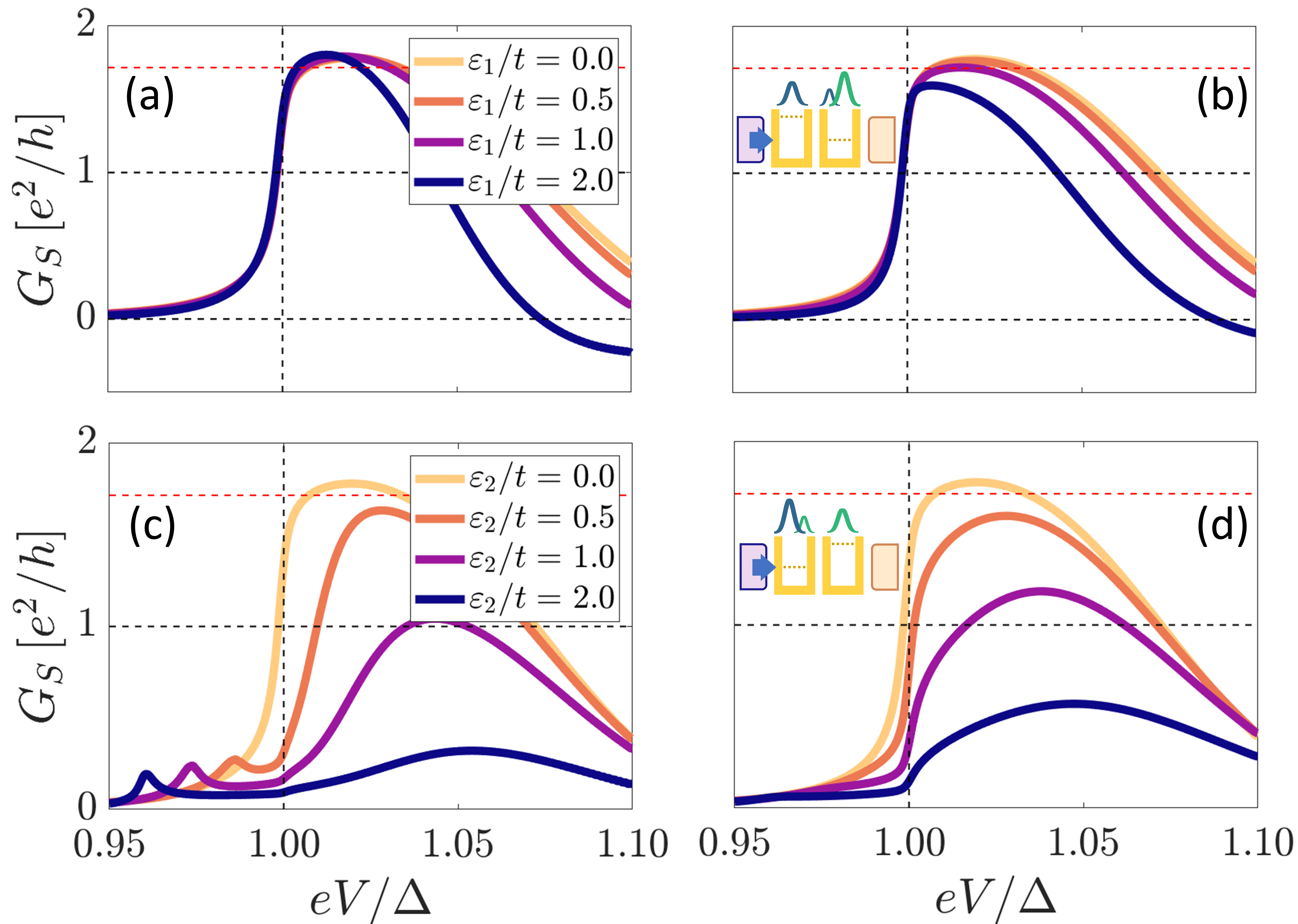}
\caption{
Local (a,b) and non-local (c,d) effects in SC transport in the weak coupling limit, varying the chemical potential of the dot in contact with the respective probe. First (Second) row shows the dependence of the SC conductance for various values of $\varepsilon_1$ ($\varepsilon_2$). The different columns show different coupling to the normal lead $\lambda_n/t = 0.1$ (a,c) and $\lambda_n/t = 0.5$ (b,d). Inset shows a diagram of the setup, where the blue arrow marks the direction of the probing. Red dashed line marks the universal height value $G_s/G_0 = (4-\pi)$. We set $\lambda_s/t = 0.5$ in all panels.}
\label{fig2}
\end{figure}

The poles of the Green's function are contained in the characteristic polynomial $P(\omega) = g_e g_h - f^2$ with
\bea \label{full_poly_al}
g_{e/h}(\omega) &=& \gamma_+ \gamma_- \left[ (\omega \pm \varepsilon_1)\tau + \omega \Gamma_s \right] -(t^2 \gamma_{\mp}+\Delta_p^2\gamma_{\pm})\tau \, , \nonumber \\
f(\omega) &=& -(\gamma_+ + \gamma_-)t\Delta_p \tau \, ,
\eea
where $\tau = \sqrt{\Delta_s^2-\omega^2}$ and $\gamma_\pm =(\omega \pm \varepsilon_2 - i\Gamma_n)$. In the sweet spot ({\it i.e.}, $t=\Delta_p$ and $\varepsilon_1 = \varepsilon_2 = 0$) and $\Gamma_n\rightarrow 0$ limit, the polynomial has 8th degree and simplifies to,
\begin{subequations}\label{poly_al}
\begin{align} 
P(\omega) &= \omega^4 \big (\Gamma_s+\sqrt{\Delta_s^2 - \omega^2} \big) \, Q(\omega) \, ,  \\
Q(\omega) &= \omega^2\Gamma_s + (\omega^2-4 \Delta_p^2)\sqrt{\Delta_s^2 - \omega^2} \, ,
\end{align}
\end{subequations}
where the poles of $Q(\omega)$ may describe the excited states ($\pm z_{s,e}$) in certain parameter regimes. Notice that the non-perturbative coupling to the SC lead renormalizes the energy of such excited states, appearing as subgap poles detaching from the continuum. These poles can be described analytically, and we reproduce here an approximation to the lowest contribution in $\Gamma_s$ 
\be \label{excited_states_AN}
z_{s,e} \approx \frac{1}{\sqrt{3}} \sqrt{8\Delta_p^2 + \Delta_s^2 - \alpha^{1/3}(\Gamma_s) - \frac{\delta^2}{\alpha^{1/3}(\Gamma_s)}} \; ,
\ee 
where $\alpha(\Gamma_s) = \big(\delta^3 - 12\sqrt{3} \, \Gamma_s \Delta_p^2 \, \delta^{3/2} \big)$, and $\delta = (4\Delta_p^2- \Delta_s^2)$, cf. App.~\ref{App_A} for details.

\begin{figure}[b!]
\centering
\includegraphics[width=\columnwidth]{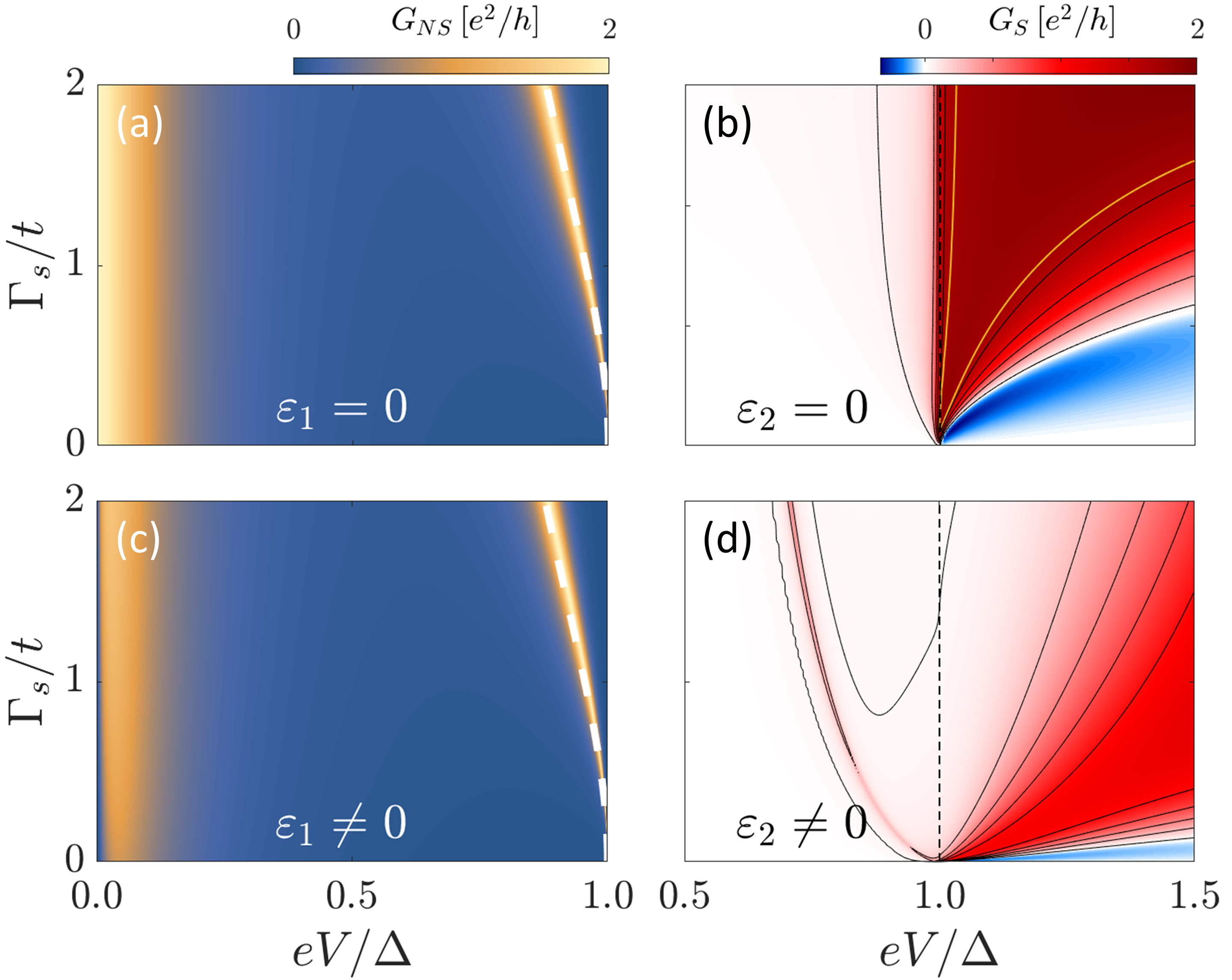}
\caption{Non-local effects in transport measurements as a function of the coupling to the SC lead. Left (Right) column, normal (SC) conductance for $\varepsilon_1/t=0$ and, $\varepsilon_1/t=1$ ($\varepsilon_2/t=0$ and, $\varepsilon_2/t=1$).  Yellow contour marks the universal height value $G_s/G_0 = (4-\pi)$. White dashed lines represents the analytic subgap pole ($z_{s,e}$) obtained from Eq.~\eqref{poly_al}. The coupling to the normal lead is set to $\lambda_n/t = 1$ (a, c), and $\lambda_n/t = 0.1$ (b, d).}
\label{fig3}
\end{figure}

\section{Weak coupling}\label{SecIII}

We first examine transport through the normal lead when the Kitaev chain is weakly coupled to the SC lead ($\Gamma_s/t \lesssim 1$), comparing it to the previously known phenomenology~\cite{Leijnse2012, Dvir2023}. 

\begin{figure*}[t!]
\centering
\includegraphics[width=1\textwidth]{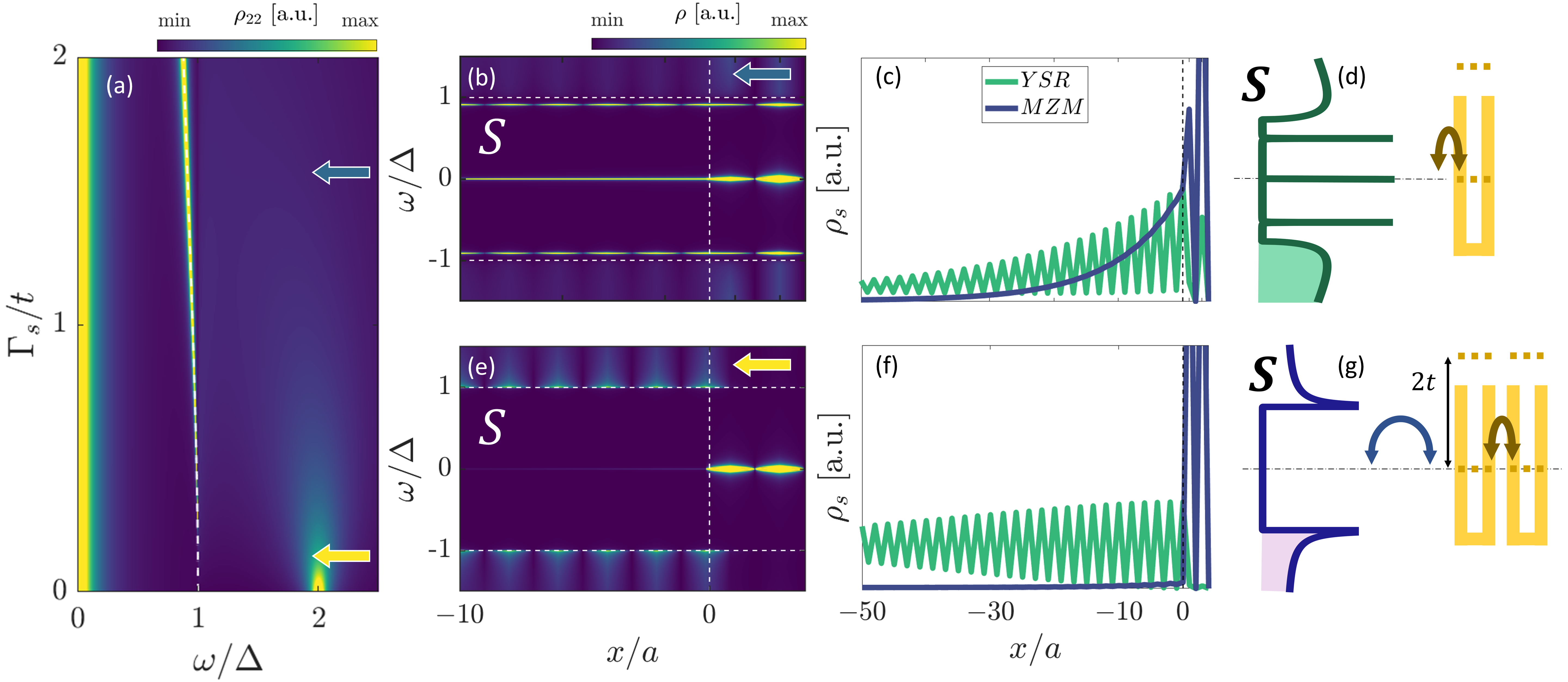}
\caption{(a) QD2 LDOS as a function of the hybridization parameter to the SC lead. (b,e) LDOS across the junction, showing the spreading of the states into the SC lead in the $\lambda_n/t \rightarrow 0$ limit. (c,f) LDOS cuts following the YSR state and the MZM respectively. (b,c) [(e,f)] Strong (Weak) coupling for $\lambda_s/t = 4$ ($\lambda_s/t = 0.5$) marked by arrows in (a). (d, g) Illustration of the weak and strong coupling across the junction (d) Weak coupling, minimal Kitaev chain. (g) Strong coupling, YSR state coupled to QD2 forming bonding and anti-bonding subgap states. White dashed line in panel (a) represents the analytic pole obtained from Eq.~\eqref{full_poly_al} when considering $\Gamma_n \neq 0$. The LDOS at QD1 and QD2 in (c,f) is rescaled for visualization purposes. We set $\lambda_n/t=1$ in the rest of the panels.}
\label{fig4}
\end{figure*}

Figure~\ref{fig1}(b) illustrates the Majorana localization when detuning the two QDs away from the sweet spot. When QD1 is detuned, the two MZMs have a finite spectral weight in QD2. Thus, when considering $\varepsilon_1 \neq 0$, the normal lead couples to both MZMs, producing a destructive interference sharp dip at zero energy, Fig.\ref{fig1}(c), as previously observed in Ref.~\cite{Leijnse2012}. Moreover, due to electrons tunneling between the normal and the SC lead, mediated by the Kitaev chain, a peak appears at $e|V|=\Delta$, where $\Delta_p=\Delta_s=\Delta$, see also Fig.\ref{fig1}(d). 

By contrast, when QD2 is detuned from the sweet spot, the two MZMs acquire a finite weight in QD1. This leads to a quantized zero-bias conductance peak, showing a width that depends on $\varepsilon_2$, cf. Fig.\ref{fig1}(f) and Ref.~\cite{Leijnse2012}. Therefore, detuning either the QD attached to the normal lead or the one attached to the SC lead has very different effects on the transport properties. 

We now systematically analyze the role of level detuning in SC transport in the regime of weak coupling to the normal lead ($\Gamma_n/t \lesssim 1$), see Fig.~\ref{fig2}. Specifically, we concentrate on the features close to the gap, as subgap transport is suppressed due to the QD spin-polarization. Figures~\ref{fig2}(a,b) show how SC transport is locally modified as the QD directly coupled to the SC lead is detuned, therefore pushing the MZM spectral weight to the opposite QD. The conductance measured in the SC exhibits an approximately quantized value, $(4-\pi)$ at $e|V|=\Delta$, almost independently of the value of $\varepsilon_1$, thus reproducing the phenomenology predicted for nanowires~\cite{Peng2015, Zazunov2016}, see Fig.\ref{fig2}(a). However, a small reduction in the conductance height at $e|V|=\Delta$ can be seen for larger values of $\Gamma_n$ and $\varepsilon_1 \gtrsim t$, see Fig.\ref{fig2}(b). 

Figure~\ref{fig2}(c,d) illustrates the opposite behaviour, namely how SC transport is modified non-locally when the QD far from the SC lead is detuned. In this case, the lead couples to both MZMs. The resulting hybridization becomes evident when testing transport through the SC probe, where the quantized conductance value gets reduced as soon as $\varepsilon_2\neq 0$. Moreover, the coherent coupling between MZMs mediated by the SC lead, gives rise to subgap signatures in transport, Fig.~\ref{fig2}(c). We remark that the subgap structure is not associated to any renormalization of subgap poles but to inelastic single-particle processes in transport, cf. Refs.~\cite{Ruby2015} and App.~\ref{App_D}. When the coupling to the normal lead is increased, the spreading of the MZM into the latter reduces the hybridization between MZMs and therefore, enhances the SC transport, Fig.~\ref{fig2}(d). Notice that this dependence with $\Gamma_n$ is the opposite to the one found in the $\varepsilon_1\neq0$ case in Fig.~\ref{fig2}(a, b).

\section{Strong coupling}\label{SecIV}

\begin{figure*}[t!]
\centering
\includegraphics[width=1\textwidth]{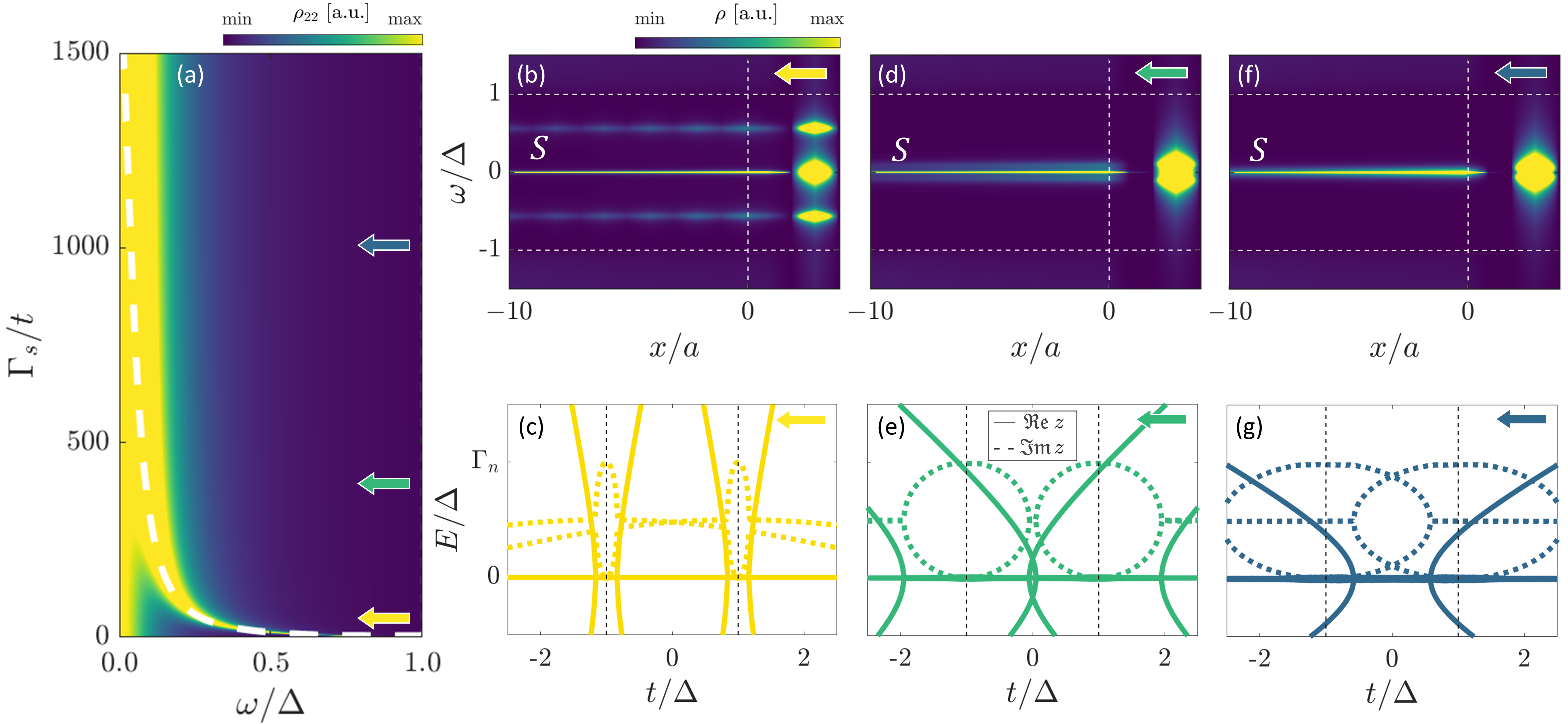}
\caption{(a) QD2 LDOS as a function of the hybridization parameter to the SC lead increasing from weak to the ultra-strong coupling limit. First row, LDOS across the junction, showing the spreading of the states into the SC lead. Second row, real and imaginary part of the poles of the system (solid and dotted lines respectively) for $\lambda_s/t = 10, 60, 100$ marked by arrows in (a). White dashed line in panel (a) represents the analytic pole obtained from Eq.~\eqref{full_poly_al} when considering $\Gamma_n \neq 0$. We set $\lambda_n/t=1$ in all panels, except indicated, we have set $t/\Delta=1$. }
\label{fig5}
\end{figure*}

Now we examine the electron transport as a function of the coupling to the SC lead, going beyond the weak coupling regime ($\Gamma_s/t \gtrsim 1$) where the excited state $z_{s,e}=2\Delta$ is strongly renormalized by the lead, emerging as a subgap YSR state \cite{Ruby2015, Aguado2017b, Scherubl2020, Cuevas2020b, Miles2023, Liu2023}. Transport provides an ideal method to observe the interplay between the aforementioned YSR states and MZMs, Fig.~\ref{fig3}. The onset of the YSR state appears as a new peak detaching from the continuum with linear dispersion around $\Gamma_s/t \gtrsim 0.5$, Figs.~\ref{fig3}(a, c). This phenomenology does not affect the MZM signatures at $V\ll\Delta$, except for $\Gamma_s/t\gg1$ and finite spin-polarization in QD1, see below and App.~\ref{App_C}. The conductance of the excited state is also quantized for $\Gamma_s/t \geq 1$ and $\varepsilon_2=0$. Intuitively, the excited state is no longer broadened by the SC lead when is well detached from the continuum, thus recovering the Majorana properties of the original excited state.

Detuning $\varepsilon_1$ barely affects the excited state and only leads to a dip in the zero-bias conductance, as commented in the weak coupling regime, see Fig.~\ref{fig1}. The energy of the excited state is renormalized when $\varepsilon_2$ shifts away from the sweet spot, losing any quantized behaviour in transport. In particular, it gets pinned to $z_{s,e} \rightarrow \varepsilon_2$ for large $\Gamma_s$ values, see Fig.~\ref{figS2} in App.~\ref{App_A}.

Transport through the SC lead is also affected when $\Gamma_s$ increases, Fig.~\ref{fig3}(b, d). First, the universal height value $(4-\pi)$ moves to larger values of the voltage bias as the coupling to the SC lead is increased~\cite{Ruby2015}, illustrated by the yellow line. Figure~\ref{fig3}(d) shows the onset of a subgap structure associated to the hybridization between Majoranas when detuning QD2 even in the weak coupling regime $\Gamma_s \rightarrow 0$, see also Fig.~\ref{fig2}(c). This subgap structure is obtained from the single-particle contribution when testing transport through the SC probe when enabling thermally activated inelastic cotunneling. Thus, the hybridized MZMs behaves as a trivial zero energy subgap state in the strong tunneling regime~\cite{Ruby2015}, cf.  App.~\ref{App_D} and \ref{App_E}. Notice that the conductance peaks have a reduced value, compared to panel (b) and for $\Gamma_s/t \leq 1$ the dispersion of the subgap peak is linear with $\lambda_s$ ({\it i.e.}, quadratic with $\Gamma_s$) highlighting the different origin with the observed subgap structure in Fig.~\ref{fig3}(a,c). 

\section{Interplay between YSR and MZMs}\label{SecV}

The previous transport characteristics can be understood in terms of the Local Density of States (LDOS) which provides a way to study the transition from the weak to strong coupling to the leads ({\it i.e.,} $\Gamma_s/t \gtrsim 1 \gtrsim \Gamma_n/t$), resolving the degree of localization of the different subgap states, cf. App.~\ref{App_B} for details. 

The LDOS at QD2 Fig.~\ref{fig4}(a), features a transition around $\Gamma_s/t\sim0.5$ in which the excited state $z_{s,e}=2\Delta$ disappears and re-appears as a subgap state. The transition also reflects on a different localization of the states in the strong and weak coupling regimes, see Figs.~\ref{fig4}(b,e) respectively. For weak coupling to the SC ($\Gamma_s/t\ll1$), the system hosts confined MZMs where the SC acts as a gapped lead, Fig.~\ref{fig4}(f). Therefore, we characterize this regime as a minimal Kitaev chain, where the excited states are just broadened by the coupling to the leads, see Fig.~\ref{fig4}(g). In contrast, in the strong coupling regime $\Gamma_s/t\gtrsim 1$, both the MZM and the subgap state at $\omega\lesssim\Delta$, extend into the SC lead, Fig.~\ref{fig4}(c). In this limit, the SC lead + QD1 system is described by a YSR-like state, where QD1 is subsumed in the SC lead, see Fig.~\ref{fig4}(d). The coupling between the YSR state and QD2 renormalizes the excited state, reducing its energy, while the MZM remains at $\omega=0$. Thus, the correct physical picture is no longer that of a minimal Kitaev chain weakly coupled to both SC and normal leads, but rather of a YSR coexisting with a MZM.

The energy of these YSR states decreases as $\Gamma_s$ increases, merging with the MZMs at zero energy for $\Gamma_s/t \rightarrow \infty$ for a fully spin-polarized QD1. A non-zero $\Gamma_n$ allows for the merging to occur at a finite value of the coupling with the SC lead, making it hard to demonstrate the Majorana properties of the chain as the gap to excited states goes to zero~\cite{Cayao2024}, see Fig.~\ref{fig5}(a). The dashed line shows the convergence of the YSR state to zero energy, described by Eq.~\eqref{poly_al}, and the width of the state is determined by $\Gamma_n$. Even though the YSR and the Majorana states overlap, the oscillatory behaviour of the YSR is lost \cite{Pal2024} and the zero-bias conductance quantization is preserved, cf. App.~\ref{App_G}. When considering $\varepsilon_1 \neq 0$ or $t/\Delta \neq 0$, the subgap states also tend towards zero energy in the ultra-strong coupling regime, however a conductance dip at zero-bias is observed associated to a destructive interference, cf. App.~\ref{App_G}. 
 
The merging of the YSR and the MZM states can be understood in terms of the LDOS across the junction and the EP associated to the zero poles, {\it i.e.}, non-Hermitian spectral degeneracies in the complex spectrum of the system when considering the strong coupling to the leads. These EPs are seen as bifurcations in the poles' imaginary part, where eigenvalues and eigenvectors coalesce~\cite{Berry2004}. For $\Gamma_s/t \geq 1$, along with the EP associated to the MZM, a new secondary EP appears linked to the quantized conductance behaviour of the YSR excited state in this regime, cf. App.~\ref{App_A}. Moreover, we observe a new regime at $\Gamma_s/t \gtrsim 4$ such there is a transition of the pole defining the dispersion of the YSR state, see also Fig.~\ref{figS1} in App.~\ref{App_A}. Furthermore, in this regime the MZM at QD1 is being absorbed by the SC lead, Fig.~\ref{fig5}(b).

For larger values of the coupling, the EPs centered around $\{-t,t\}$ start to touch each other, signalling the overlap of the subgap states, see Fig.~\ref{fig5}(c,e). The spectral weight at zero energy in QD1 is suppressed as the MZM is delocalized into the lead, Fig.~\ref{fig5}(d). For even larger values of the coupling, the EPs intersects and the extended MZM living in the SC lead becomes narrower and narrower, Fig.~\ref{fig5}(f,g). Thus, we observe a non-destructive overlap between the YSR and MZMs that preserve a confined zero energy subgap state with quantized conductance.

\begin{figure}[t!]
\centering
\includegraphics[width=1\columnwidth]{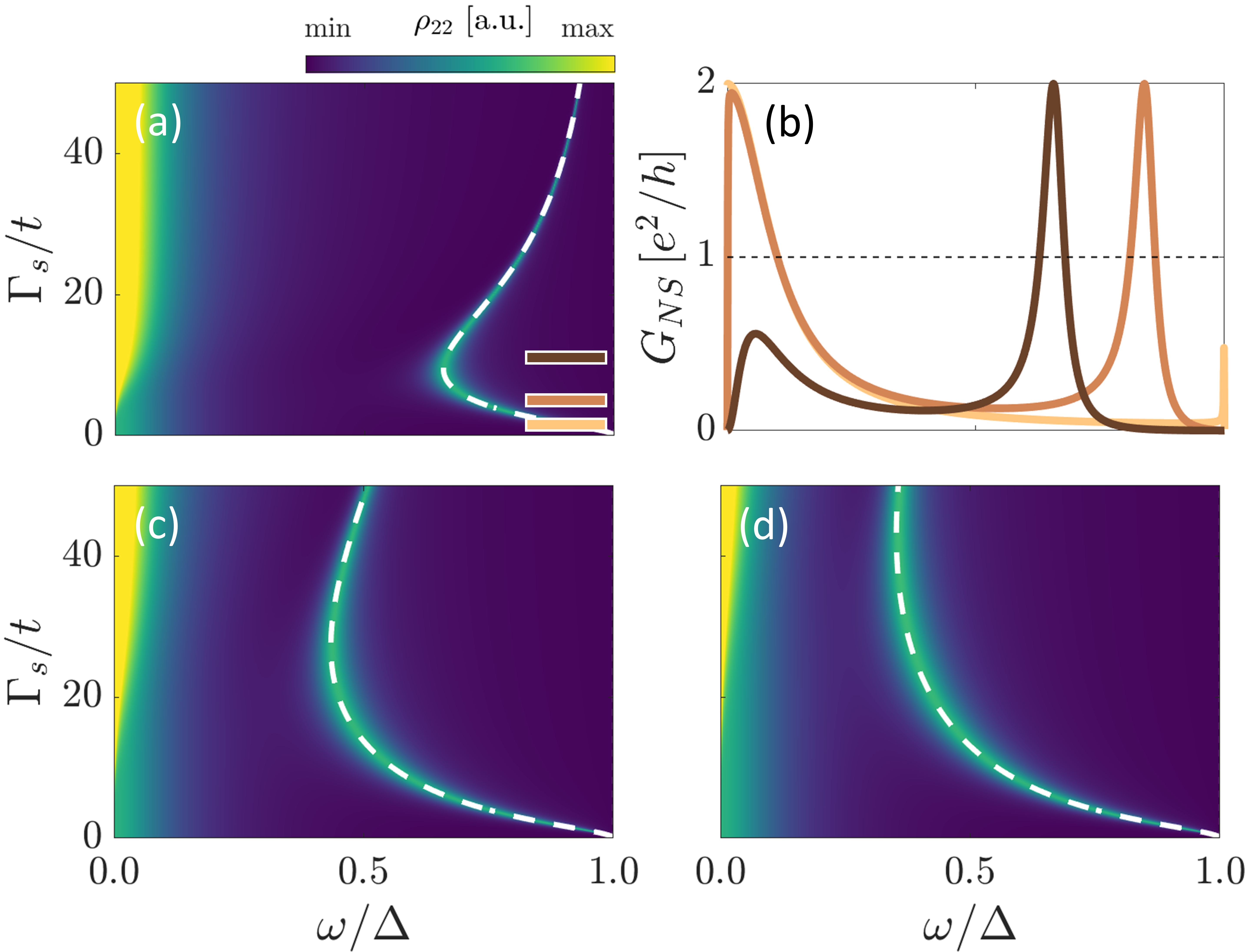}
\caption{Interplay between YSRs and MZMs when partial spin-polarization is considered in QD1. (a) QD2 LDOS as a function of the hybridization parameter to the SC lead going to the ultra-strong coupling for $E_z/t = 10$. (b) Normal conductance for $\lambda_s/t = 1, 5, 10$ marked in (a). (c,d) QD2 LDOS for $E_z/t = 50, 100$. White dashed line represents the computational pole obtained from the characteristic polynomial when considering $\Gamma_n, E_z \neq 0$. We set $\lambda_n/t=1$ in all panels.
} 
\label{fig6}
\end{figure}

\section{Finite spin-polarization}\label{SecVI}

In real experiments, the QDs have a finite spin polarization. Despite both QDs being subjected to the same global magnetic field, a strong coupling to the SC will strongly renormalize the g-factor of QD1~\cite{Miles2023}. To account for this effect, we consider a finite Zeeman splitting in QD1 while keeping QD2 spin-polarized ({\it i.e.}, $E_{z,1} = E_z$, and $E_{z,2}\rightarrow \infty$), although our findings remain qualitatively invariant for a finite spin-splitting in QD2.

Due to the partial polarization in QD1, the coupling with the SC lead is able to mix spin up and down, leading to a competition between the Zeeman splitting and the hybridization with the SC. This competition reflects in the density of states as an avoided crossing between the MZMs and the YSR states, see Fig.~\ref{fig6}(a). After the maximum approaching, the Majorana properties are spoiled, including the quantized zero-bias conductance, see Fig.~\ref{fig6}(b). The transition happens for $\Gamma_s \sim E_Z$. Moreover, when the YSR state is well detached from the continuum, it exhibits quantized conductance behaviour and coupling dependent width. However, this quantized regime is reached when the ZBP is completely lost, as a signature of the destructive interference in contrast to the fully spin-polarized case, cf. App.~\ref{App_C} for details on the topological analysis in this regime.

Figure~\ref{fig6}(c,d) reflect that as the Zeeman field in QD1 increases, the crossing point is displaced to larger values of the coupling to the SC lead in such a way that for full polarization, the regime of destructive interference is never reached and the gap to the excited state reduces with $\Gamma_s$~\cite{Liu2023}. Therefore, in the ultra-strong coupling regime, the degree of polarization is relevant to the physics of the problem. In this regime, MZMs are present only for $\Gamma_s \ll E_z$. Thus, it is important to preserve a well-defined spin-polarization of the YSR state along with including the effects of the continuum of quasiparticles when considering Kitaev chains formed by ABS-QDs building blocks~\cite{Miles2023, Liu2023}.

\section{Conclusions and Outlook}\label{SecVII}

Artificial Kitaev chains made out of quantum dots (QDs) coupled to superconductors (SCs) have emerged as promising platforms to probe Majorana physics. In this work, we have demonstrated that a SC can be used to probe the properties of Majorana states appearing in minimal Kitaev chains, offering insights about their localization. This reflects in non-local subgap features in the SC conductance when the chemical potential of the QD far from the SC lead is detuned. 

We have characterized the weak and strong coupling to the SC lead as the evolution from a minimal Kitaev chain, to a system defined by the coexistence of a YSR state and MZMs. The energy, and therefore the gap to the excited state, decreases with the coupling to the SC. For finite spin-polarization in the QDs, the coupling to the SC induces a YSR-Majorana hybridization, spoiling the characteristic properties of the Majorana states, including the conductance quantization. This illustrates the competition between YSR and Majorana physics in minimal Kitaev chains.

We foresee the extension of the analysis to hybrid setups where one could induce transport through an Andreev bound states, {\it e.g.} by including an extra QD between the SC lead and the minimal Kitaev chain, leading to extra resolution of subgap features. 

\acknowledgments
We thank A. Bordin and M. Pino for valuable comments and discussions. 
Work supported by the Horizon Europe Framework Program of the European Commission through the European Innovation Council Pathfinder Grant No. 101115315 (QuKiT), the Spanish Comunidad de Madrid (CM) “Talento Program” (Project No. 2022-T1/IND-24070), the Spanish Ministry of Science, innovation, and Universities through Grants PID2022-140552NA-I00,  PID2021-125343NB-I00 and TED2021-130292B-C43 funded by MCIN/AEI/10.13039/501100011033, “ERDF A way of
making Europe” and European Union Next Generation
EU/PRTR. Support from the CSIC Interdisciplinary Thematic
Platform (PTI+) on Quantum Technologies (PTI-QTEP+) is also acknowledged.

\appendix

\setcounter{equation}{0}
\renewcommand{\theequation}{S\,\arabic{equation}}

\setcounter{figure}{0}
\renewcommand{\thefigure}{S\,\arabic{figure}}

\section{Poles Structure and EPs}\label{App_A}

In the main text, we have described a hybrid setup consisting in a quantum dot (QD) based minimal Kitaev chain coupled to a normal and a superconducting (SC) lead which can be described by the advanced boundary Green's functions (bGF)~\cite{Cuevas1996, Zazunov2016, Alvarado2020, Alvarado2021, Alvarado2022} written 
in the $4\times4$ (hat notation) Nambu basis $\hat{\Psi} = (\psi_\up, \psi_\up^\dagger, \psi_\dw, \psi_\dw^\dagger)^T$,
\be
\hat{{\cal G}}^A_n = \frac{i}{t_n}\sigma_0\tau_0 \ , \quad \hat{{\cal G}}^A_s = -\frac{\omega \sigma_0\tau_0 - \Delta_s\sigma_y\tau_y}{t_s \sqrt{\Delta_s^2-\omega^2}} \ ,
\ee 
where $\tau_\mu$ ($\sigma_\mu$) are Pauli matrices acting in spin (electron/hole) space, both in the wide-band approximation $t_n,\, t_s \gg \omega, \,\Delta_s$. For the advanced GFs we are implicitly assuming $\omega \rightarrow \omega - i\eta$ with $\eta \rightarrow 0^+$. Following Ref.~\cite{Zazunov2016} we define the brunch cut in the square root along the negative axis,
\be
\sqrt{\Delta_s^2-(\omega-i\eta)^2}= \left\lbrace
\begin{array}{cl}
\sqrt{\Delta_s^2-\omega^2} \, , & |\omega|\leq \Delta_s \, , \\
i \, \textrm{sgn}(\omega)\sqrt{\omega^2-\Delta_s^2} \, , & |\omega| > \Delta_s \, .
\end{array}
\right.
\ee
From now on, we omit the super-index $A$.

The coupling terms between the different elements across the junction take the form $\hat{\Sigma}_{n/s} = \lambda_{n/s} \sigma_z\tau_0$ and, the toy model for the minimal Kitaev chain~\cite{Leijnse2012} written in this basis,
\be \label{Hamiltonian}
\hat{H}_{\nu} = \varepsilon_{\nu} \sigma_z \tau_0 + 2E_{z,\nu} \sigma_z \Pi_{\dw}  \ , \quad \hat{T} = (t\sigma_z + i\Delta_p \sigma_y)\Pi_{\up} \ ,
\ee 
where $\nu={1,2}$ corresponds to the different sites in the chain, and $\Pi_{\up, \dw} = (\tau_0 \pm \tau_z)/2$ are the projectors over the spin up/down sector. Notice that as one reaches the polarized limit of the minimal Kitaev chain when $E_{z,\nu}\rightarrow\infty$, the coupling between the QDs in the spin down sector goes correspondingly to zero. It should be reminded that the effective coupling between QDs is obtained from a Schrieffer–Wolff transformation and therefore, it is inversely proportional to the energy difference between states. Thus, the GFs projected over the minimal Kitaev chain can be computed  working directly on a spinless $2\times2$ basis $\Psi = (\psi_\up, \psi_\up^\dagger)^T$, equivalent to remove the anomalous part of the SC lead's GF in the calculation. This effect is associated to the incompatibility between order parameters between the SC lead and the Kitaev chain~\cite{Zazunov2012, Peng2015, Zazunov2016}. Unless otherwise stated, we use $\Delta_p=\Delta_s=\Delta=t$ and $t_s=t_n=10 \, t$.
In the spin-less basis it is satisfied that
\be
{\cal G}_n = \frac{i}{t_n}\sigma_0 \ , \quad {\cal G}_s = -\frac{\omega \sigma_0}{t_s \sqrt{\Delta_s^2-\omega^2}} \ ,
\ee 
and the total GF projected over the QDs takes the form
\begin{subequations}\label{totalGF}
\begin{align}
G_{11} ={}& \big[[{\cal G}_{11}]^{-1} - \Sigma_s^\dagger{\cal G}_s\Sigma_s \big]^{-1} ,  \\
G_{22} ={}& \big[[{\cal G}_{22}]^{-1} - \Sigma_n{\cal G}_n\Sigma_n^\dagger \big]^{-1} ,
\end{align}
\end{subequations}
where the bGF associated to both ends of the minimal Kitaev - including both QD in the chain and the opposite end lead - are
\begin{subequations}
\begin{align}
{\cal G}_{11} ={}& [\omega \mathbb{I} - H_1 - T {\cal G}^{(0)}_{22} T^\dagger]^{-1} ,  \\
{\cal G}_{22} ={}& [\omega \mathbb{I} - H_2 - T^\dagger {\cal G}^{(0)}_{11} T]^{-1} ,
\end{align}
\end{subequations}
and 
\begin{subequations}
\begin{align}
{\cal G}^{(0)}_{11} ={}& [\omega \mathbb{I} - H_1 - \Sigma_s^\dagger{\cal G}_s\Sigma_s]^{-1} ,  \\
{\cal G}^{(0)}_{22} ={}& [\omega \mathbb{I} - H_2 - \Sigma_n{\cal G}_n\Sigma_n^\dagger]^{-1},
\end{align}
\end{subequations}
being the bGF of each QD in the chain coupled to its adjacent lead.

Without any loss of generality, the poles of the system (shared by both QDs) can be obtained from $G_{11}$ in Eq.~\eqref{totalGF}. Then, the associated self energy due to the coupling to QD2 including the adjacent normal lead $\Sigma = T \, {\cal G}^{(0)}_{22} T^\dagger$, simplifies to
\be 
\Sigma =  
\bmat t^2 \Omega_- + \Delta_p^2 \Omega_+ && -t\Delta_p (\Omega_+ + \Omega_-) \\ -t \Delta_p (\Omega_+ + \Omega_-) && t^2 \Omega_+ + \Delta_p^2 \Omega_- \emat,
\ee 
where $\Omega_{\pm} =1/(\omega \pm \varepsilon_2 - i\Gamma_n) = 1/\gamma_\pm$ describes the poles associated to the QD2 coupled to the normal lead, being $\Gamma_n = \lambda_n^2/t_n$ the hybridization parameter with the normal lead.

\begin{figure*}[t!]
\centering
\includegraphics[width=0.8\textwidth]{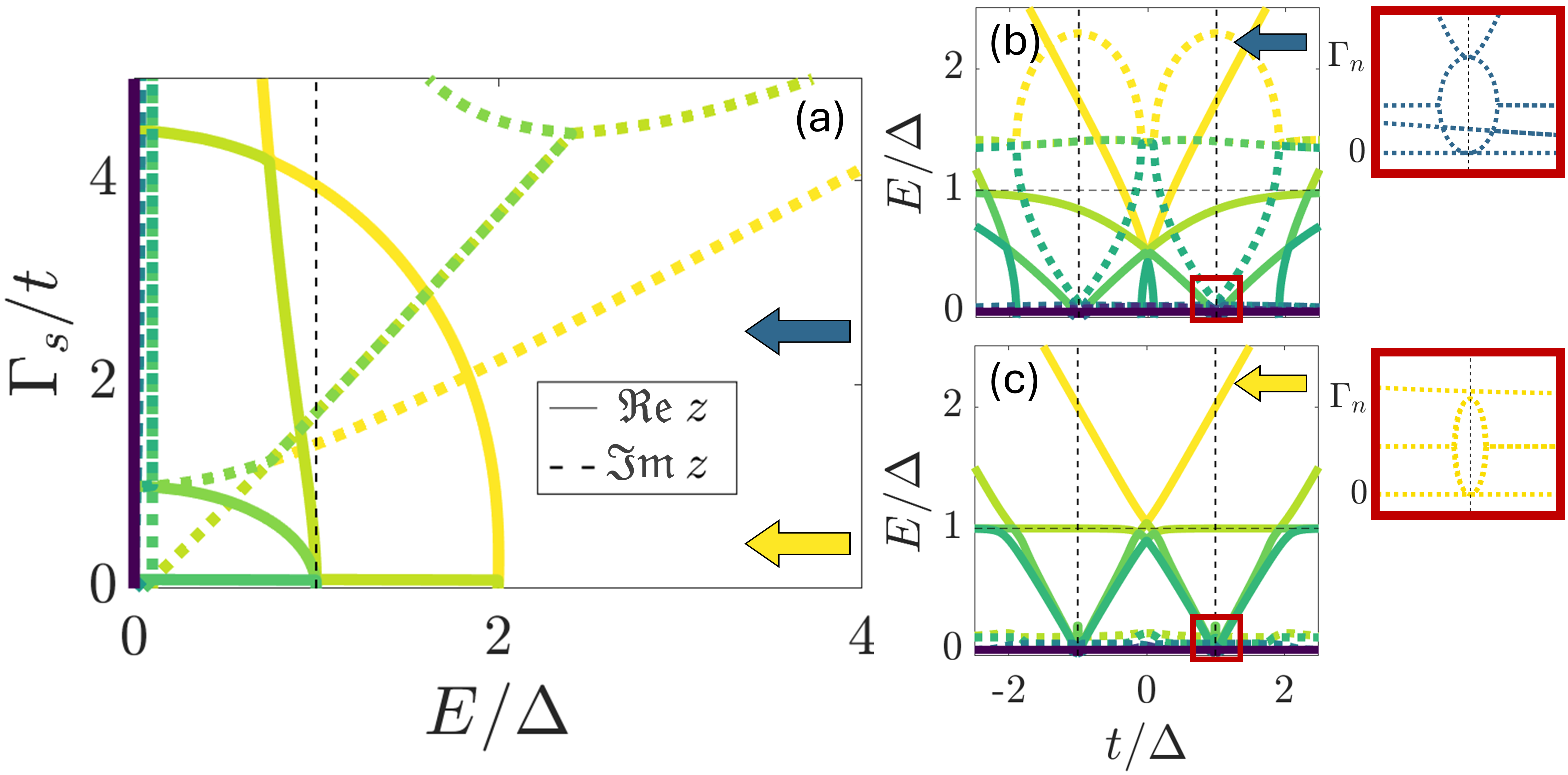}
\caption{(a) Real and imaginary part (solid and dotted lines respectively) of the poles as a function of the hybridization parameter to the SC lead obtained from Eq.~\eqref{poly_al_sm} when considering $\Gamma_n \neq 0$. (b, c) EPs for $\lambda_s/t = 5$ ($\lambda_s/t = 2$) marked by arrows in (a). Insets show a close-up image of the fundamental EPs. We set $\lambda_n/t=1$ in all panels, except indicated, we have set $t/\Delta=1$.} 
\label{figS1}
\end{figure*}

Finally, the analytical expression for $G_{11}$ can be written as
\be \label{G11_an}
G_{11} = \frac{\tau \, \gamma_+ \gamma_- }{P(\omega)}\bmat g_e(\omega) && f(\omega) \\ f(\omega) && g_h(\omega) \emat ,
\ee
where
\bea
g_{e/h}(\omega) &=& \gamma_+ \gamma_- \left[ (\omega \pm \varepsilon_1)\tau + \omega \Gamma_s \right] -(t^2 \gamma_{\mp}+\Delta_p^2\gamma_{\pm})\tau \, , \nonumber \\
f(\omega) &=& -(\gamma_+ + \gamma_-)t\Delta_p \tau \, ,
\eea
and $\tau = \sqrt{\Delta_s^2-\omega^2}$, where $\Gamma_s = \lambda_s^2/t_s$ is the hybridization with the SC lead. Notice that the characteristic polynomial $P(\omega) = g_e g_h - f^2$ has 8th degree in $\omega$ and contains all the information concerning the poles of the system. In the sweet spot and $\Gamma_n\rightarrow 0$ limit, the polynomial simplifies to
\begin{widetext}
\be \label{poly_al_sm}
P(\omega) = \omega^4 \big (\Gamma_s+\sqrt{\Delta_s^2 - \omega^2} \big) \left[\omega^2\Gamma_s +  (\omega^2-4 \Delta_p^2)\sqrt{\Delta_s^2 - \omega^2} \right].
\ee
Along with the trivial zeros and $z_0 = \sqrt{\Delta_s^2-\Gamma_s^2}$, the rest of the poles can be computed analytically as,
\begin{subequations}
\begin{align} \label{YSR_pole}
z_\mu =& \, \frac{1}{\sqrt{6}} \sqrt{2 (8 \Delta_p^2 + \Delta_s^2 - \Gamma_s^2) + \xi_{\mu}^* \alpha^{1/3}(\Gamma_s) + \xi_{\mu} \frac{ (4 \Delta_p^2 - \Delta_s^2)^2 - 2 \Gamma_s^2 (8 \Delta_p^2 + \Delta_s^2) +  \Gamma_s^4 }{\alpha^{1/3}(\Gamma_s)}   } \; .  \\ \nonumber \\
\alpha(\Gamma_s) =& \, (4 \Delta_p^2 - \Delta_s^2)^3 + 3 \Gamma_s^2 (40 \Delta_p^4 + 4 \Delta_p^2 \Delta_s^2 + \Delta_s^4) - 3 \Gamma_s^4 (8 \Delta_p^2 + \Delta_s^2) + \Gamma_s^6 - 
  12 \sqrt{3} \, \beta(\Gamma_s) \, , \\
\beta(\Gamma_s) =& \, \Gamma_s \Delta_p^2\sqrt{ (4 \Delta_p^2 - \Delta_s^2)^3 - 2\Gamma_s^2(2 \Delta_p^4 -10 \Delta_p^2 \Delta_s^2 - \Delta_s^4)  - \Gamma_s^4 \Delta_s^2  } \, ,
\end{align}
\end{subequations}
\end{widetext}
where the parameter $\xi_{\mu}$ can be either $\xi_1=-2$ or $\xi_2=(1 + i \sqrt{3})$. The subset of non-trivial poles of the system are contained in $\{z_0, z_1, z_2 \}$, along with their symmetry partners $z \rightarrow -z^*$. It should be noted that by the choice of advanced GFs, all the physical poles satisfy $\Im z \geq 0$. Depending on the parameter regime, the excited state of the system ($z_{s,e}$) may be described by one of these subgap poles. In the main text we reproduce the lowest order in the superconducting hybridization of $z_1$ to give a comprehensible compact approximation of the excited state, {\it i.e.,} $\alpha \rightarrow (4 \Delta_p^2 - \Delta_s^2)^3 - 12 \sqrt{3} \,\beta$, and $\beta \rightarrow  \Gamma_s \Delta_p^2 (4 \Delta_p^2 - \Delta_s^2)^{3/2}$.

Figure~\ref{figS1} shows the real and imaginary parts of the poles of the system \cite{Nagai2020}, including non-Hermitian effects such $\Gamma_n\neq0$. Thus, we are able to study the Exceptional Points (EPs)~\cite{SanJose2016, Avila2019, Nagai2020, Cayao2024, Pino2024, Cayao2024b} for different values of the coupling to the SC lead. Figure~\ref{figS1}(a) shows several transitions where the real part of some poles goes to zero. First, a transition at $\Gamma_s/t=1$ appears as one of the poles detaching from the continuum $E=\Delta$, goes to zero ({\it i.e.}, $\Re z_0 \rightarrow 0$), defining the change of regime  from weak to strong coupling. 

In the weak coupling limit, there are only EPs associated to the Majorana Zero Modes (MZMs), where the excited state can be found at $z_{s,e}= 2\Delta$. Figure~\ref{figS1}(c) illustrates the EP associated to the imaginary part of the zero poles in this regime. The maximum (minimum) value of the bifurcation at the sweet spot is $\Gamma^+_1 = \Gamma_n$ ($\Gamma^-_1=0$) signaling a configuration with maximum coupling asymmetry~\cite{Avila2019}, {\it i.e.}, one of the MZM is non-decaying and completely decoupled from the SC lead~\cite{SanJose2016, Avila2019, Pino2024} owing to the SC gap in the weak coupling. 

Above the transition, the excited state behaves as a Yu-Shiba-Rusinov (YSR) state defined by the subgap pole $z_{s,e}= z_1$ in Eq.~\eqref{YSR_pole}. Remarkably, in this regime appears other secondary EP coexisting with the main one, see Fig.~\ref{figS1}(b). This secondary EP can be related to the quantized transport behaviour of the YSR in the normal conductance mentioned in the main text. The maximum (minimum) value of the bifurcation at the sweet spot for the secondary EP is $\Gamma^+_2 \approxeq \Gamma_s$ ($\Gamma^-_2=\Gamma_n$). 

Finally, Fig.~\ref{figS1}(a) shows another transition when $\Re z_1 \rightarrow 0$ around $\Gamma_s/t\sim4$, where a bifurcation of the imaginary parts of the poles take place, defining the ultra-strong coupling regime. Above this transition, the excited state is described by $z_{s,e} =  z_2$ in Eq.~\eqref{YSR_pole}. Finally, in the limit $\Gamma_s \rightarrow \infty$, the YSR state merges with the MZMs. 

\begin{figure*}[t!]
\centering
\includegraphics[width=0.8\textwidth]{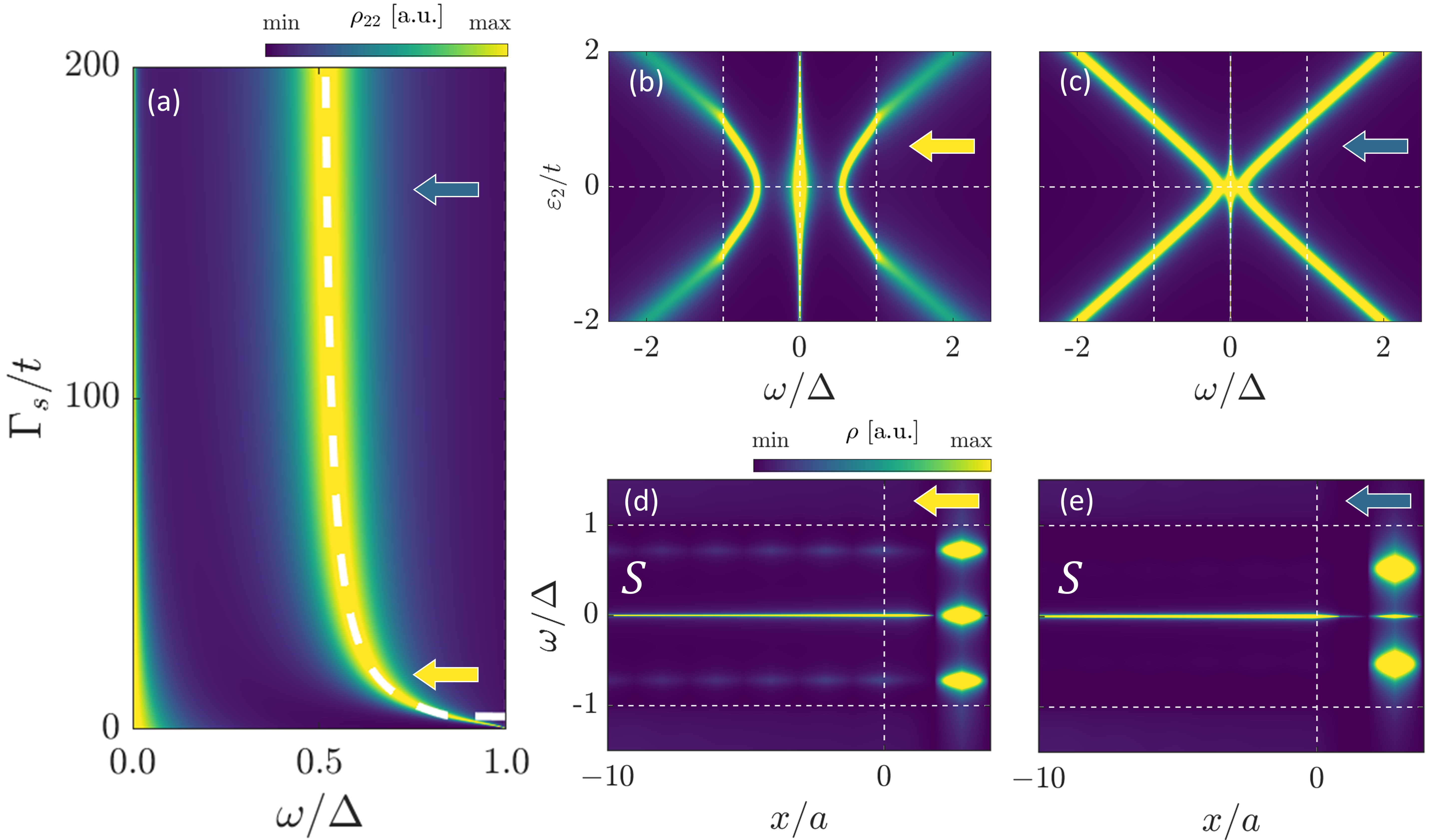}
\caption{(a) QD2 LDOS as a function of the hybridization parameter to the SC lead going to the ultra-strong coupling for $\varepsilon_2/t = 0.5$. (b-c) LDOS as a function of QD2 level for $\lambda_s/t = 10, 40$ marked by arrows in (a). (d-e) LDOS across the junction, showing the spreading of the states into the SC lead. White dashed lines represents the pole obtained computationally from the characteristic polynomial in Eq.~\eqref{G11_an}. We set $\lambda_n/t=1$ in all panels.} 
\label{figS2}
\end{figure*}

Figure~\ref{figS2} illustrates the strongly renormalized physics when detuning QD2. The excited states obey that when $\Gamma_s/t \rightarrow \infty$ then $z_{s,e} \rightarrow \varepsilon_2$. More in particular, when $\varepsilon_2 < \Delta$, the excited state is defined by a renormalized YSR state that do not merge with the MZMs, see Fig.~\ref{figS2}(a). Fig.~\ref{figS2}(d,e) shows that in the ultra strong coupling regime, the excited subgap state becomes strongly localized at QD2. Moreover, there is no longer secondary EP for large couplings (not shown), which can be related to the lack of a quantized transport response, leading to a vanishing normal transport signal in the ultra-strong coupling regime, cf. App.~\ref{App_G}. By contrast, when $\varepsilon_2 \geq \Delta$, the excited state becomes pinned to $z_{s,e} = \varepsilon_2$ even at the strong coupling transition $\Gamma_s/t>1$.

Analyzing the dispersion of the excited state with $\varepsilon_2$, for moderate couplings the excited state follows a parabolic dispersion, Fig.~\ref{figS2}(b). However, as the coupling with the SC lead is increased, the minimum of the parabola is displaced to smaller energy values and the dispersion is slightly renormalized. In the limit $\Gamma_s \rightarrow \infty$, the excited state becomes linear with $\varepsilon_2$, and the energy minimum finally collapses to zero, see Fig.~\ref{figS2}(c). By contrast, when detuning QD1, the parabolic dispersion of the excited state with $\varepsilon_1$ becomes constant, as QD1 is being absorbed by the SC lead for large couplings (not shown).

\section{Penetration of Subgap States}\label{App_B}

We follow Ref.~\cite{Alvarado2021} to analyze the penetration of the subgap poles into the leads from the total advanced GF to compute the Local Density of States (LDOS) as $\rho(\omega)=\Im \hat{G}_{\bar{j}\bar{j}}/\pi$ in the $4\times4$ Nambu basis (hat notation), where the GF in the semi-infinite SC lead acquires the form
\be 
\hat{G}_{\bar{j}\bar{j}} = \hat{{\cal G}}_{\bar{j}\bar{j}} + \hat{{\cal G}}_{\bar{j}0}  \hat{\Sigma}_s [\hat{\mathbb{I}} - \hat{{\cal G}}_{11}  \hat{\Sigma}_s^\dagger \hat{{\cal G}}_s  \hat{\Sigma}_s]^{-1} \hat{{\cal G}}_{11}  \hat{\Sigma}_s^\dagger \hat{{\cal G}}_{0\bar{j}},
\ee 
where $\bar{j} = -j$ is an integer position index running into the left SC lead and $\hat{T}_s = t_s\sigma_z \tau_0$ being the coupling between neighbouring sites inside the SC lead. The bGFs associated to the isolated semi-infinite SC lead takes the expression
\begin{subequations}
\begin{align}
\hat{{\cal G}}_{\bar{j}\bar{j}} =& \, \hat{G}^{(0)}_{00} - \hat{{\cal G}}_s \hat{T}_s \hat{G}^{(0)}_{\bar{j}0} \big[\hat{G}^{(0)}_{00} \big]^{-1} \hat{G}^{(0)}_{0\bar{j}} \hat{T}_s^\dagger \hat{{\cal G}}_s \ ,  \\
\hat{{\cal G}}_{\bar{j}0} =& \, \hat{G}^{(0)}_{\bar{j}0} - \hat{{\cal G}} _s \hat{T}_s \hat{G}^{(0)}_{\bar{j}0} \big[\hat{G}^{(0)}_{00} \big]^{-1} \hat{G}^{(0)}_{10},  \\
\hat{{\cal G}}_{0\bar{j}} =& \, \hat{G}^{(0)}_{0\bar{j}} - \hat{G}^{(0)}_{01} \big[\hat{G}^{(0)}_{00} \big]^{-1} \hat{G}^{(0)}_{0\bar{j}} \hat{T}_s^\dagger \hat{{\cal G}}_s \, .
\end{align}
\end{subequations}
Finally, the translationally invariant GFs associated to the SC bulk are defined by

\begin{subequations}
\begin{align}
\hat{G}^{(0)}_{00} =& \, [\hat{\mathbb{I}} - \hat{{\cal G}}_s \hat{T}_s \hat{{\cal G}}_s \hat{T}_s^\dagger]^{-1} \hat{{\cal G}}_s \ , \\ 
\hat{G}^{(0)}_{\bar{j}0} =& \, [\hat{{\cal G}}_s \hat{T}_s]^j \hat{G}^{(0)}_{00} \, ,  \\
\hat{G}^{(0)}_{0\bar{j}} =& \, \hat{G}^{(0)}_{00} [\hat{T}_s^\dagger \hat{{\cal G}}_s]^j \, .
\end{align}
\end{subequations}
It should be noticed that to obtain a proper convergence of the GFs across the SC lead, the bGF of the latter is defined out of the wide-band approximation, {\it i.e.}, the normal and anomalous part of the bGF, $\hat{{\cal G}}_s = g(\omega) \, \sigma_0\tau_0 - f(\omega) \, \sigma_y\tau_y$, are
\begin{subequations}
\begin{align} \label{GFs}
g(\omega) =& \, \frac{\omega}{2 t_s^2} \left(1-\sqrt{\frac{4t_s^2+\Delta_s^2-\omega^2}{\Delta_s^2-\omega^2}} \right), \\
f(\omega) =& \, \frac{\Delta_s}{2 t_s^2} \left(1-\sqrt{\frac{4t_s^2+\Delta_s^2-\omega^2}{\Delta_s^2-\omega^2}} \right).
\end{align} 
\end{subequations} 
It is trivial to see that, to compute the penetration into the normal lead, one has to transform the previous expressions following $\bar{j} \rightarrow j$ where $\hat{\Sigma}_s \rightarrow \hat{\Sigma}_n^\dagger$, $\hat{{\cal G}}_{11} \rightarrow \hat{{\cal G}}_{22}$ , $\hat{{\cal G}}_{s} \rightarrow \hat{{\cal G}}_{n}$ and $\hat{T}_s \rightarrow \hat{T}_n^\dagger = t^*_n\sigma_z\tau_0$. The normal bGF can be obtained from Eq.~\eqref{GFs} considering $\Delta_s=0$, where $\hat{{\cal G}}_{n} = (x+i\sqrt{1-x^2})/t_n \, \sigma_0\tau_0$, and $x=\omega/(2t_n)$. Notice that the analyticity of the advanced GFs imposes a change of sign of the square root for bGF of the normal lead  with respect to the SC case.

\section{Finite Spin-Polarization on QD1}\label{App_C}

\begin{figure*}[t!]
\centering
\includegraphics[width=1\textwidth]{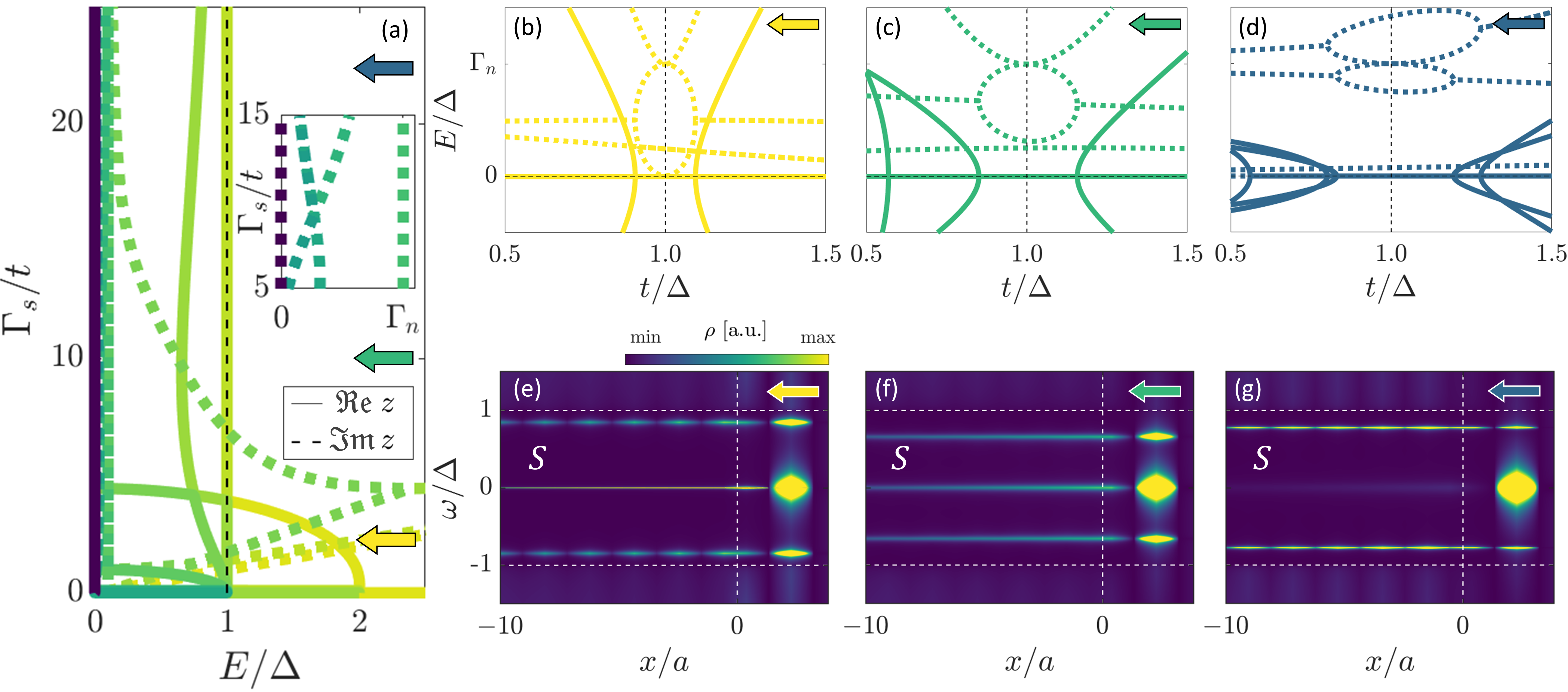}
\caption{(a) Real and imaginary part (solid and dotted lines respectively) of the poles as a function of the hybridization parameter to the SC lead obtained when considering a finite polarization in QD1. (b-d) EPs for $\lambda_s/t = 5, 10, 15$ marked by arrows in (a). (e-g) LDOS across the junction, showing the spreading of the states into the SC lead. White dashed lines represents the pole obtained computationally from the characteristic polynomial in Eq.~\eqref{G11_an} when considering a finite polarization. Inset shows a close-up image around $\Gamma_s \sim E_z$ of the imaginary part of the poles. We set $\lambda_n/t=1$ and $E_z/t = 10$ in all panels, except indicated, we have set $t/\Delta=1$.} 
\label{figS3}
\end{figure*}

In the main text, we have discussed a configuration where QD1 acquires a finite spin polarization, leading to the coexistence of the YSR and MZMs occurring at much smaller couplings with the SC lead. After the YSR minima of energy, a destructive interference occurs at $\Gamma_s \propto E_z$, after which the quantized zero-bias peak in the normal conductance disappears. To analyze this configuration, we have to consider that there are different Zeeman fields for each QD in Eq.~\eqref{Hamiltonian}, where $E_{z,1} = E_z$ is finite and, $E_{z,2} \rightarrow \infty$. Notice that the GF describing QD1 has to be described in the full $4 \times 4$ Nambu structure.

Figure~\ref{figS3} shows the poles of the system with partial polarization for different values of the coupling to the SC lead. We focus now on the negative part of the poles as it appears a new transition at $\Gamma_s \sim E_z$, where a bifurcation of imaginary parts closes and reopens, see Fig.~\ref{figS3}(a). Before the transition, the system shows a similar EPs structure as in the fully polarized case where the maximum (minimum) value of the secondary bifurcation is $\Gamma^+_2 \approxeq \Gamma_s$ ($\Gamma^-_2 = \Gamma_n$), see Fig.~\ref{figS3}(b). However, it shows non-zero values of $\Gamma^-_1 \gtrapprox 0$ for the main EP. This distorted EP structure for the finite spin polarization case could explain the lack of YSR quantized conductance signal for this regime in contrast to the fully polarized case.

At the transition, the new regime induces a destructive interference between the YSR and the MZMs, see Fig.~\ref{figS3}(c). Moreover, the non-zero values of $\Gamma^-_1 > 0$ for the main EP increases and, for the secondary EP, $\Gamma^+_2$ decreases, satisfying that $\Gamma_s \gg \Gamma^+_2 > \Gamma_n$. Above the transition, Fig.~\ref{figS3}(d) shows a strong renormalization of the EPs such $\Gamma^+_2, \, \Gamma^-_1 \rightarrow \Gamma_n$ as both of them are collapsing to the same point, signalling the spoiling of the Majorana properties after the crossing point.

Concerning the degree of localization of the subgap states across the junction, Fig.\ref{figS3}(e) shows the coexistence of the YSR and the MZMs for small couplings, where the Majorana living in QD1 is strongly delocalized into the SC lead. At the maximum approach of both subgap states $\Gamma_s \sim E_z$ in Fig.~\ref{figS3}(f), the YSR loses the oscillatory behaviour \cite{Pal2024}. For even larger values of the coupling, one finds the destructive interference regime where the MZMs living in QD1 disappears, giving as a result a trivial zero pole in QD2 and a YSR state with quantized conductance. 

\section{Superconducting Transport}\label{App_D}

For computing transport properties in junctions, we start from the general expression for the mean value of the current using Keldysh Green's functions in Refs.~\cite{Cuevas1996, Zazunov2016}. When considering the current through different SCs, it is convenient to define a Gauge in which the chemical potential difference appears as a time-dependent tunnel coupling, where the SC phase difference evolves as $\phi(t)=\phi_0 + (2eV/\hbar)t$, cf. Ref.~\cite{Zazunov2016}. By a suitable choice of basis, the coupling between the SC probe and a fully polarized Kitaev chain only involves one spin species, equivalent to remove the anomalous part of the SC lead as no multiple Andreev reflections are allowed~\cite{Zazunov2012, Peng2015, Zazunov2016} ({\it i.e.}, the probe act as a normal lead with a gap). Besides, there is still resonant Andreev transport in the minimal Kitaev chain due to p-wave correlations. Thus, the current through a BCS probe and a spinless p-wave SC in the $2\times2$ Nambu basis acquires the form, 
\be
I_S = -\frac{e}{h} \Re \int \frac{1}{2}\textrm{tr}_N \left \{ \tilde{\Sigma}^RG^K_{RR} + \tilde{\Sigma}^KG^A_{RR} \right \}d\omega \, ,
\ee 
where 
\begin{subequations}
\begin{align}
G_{RR}(\omega) ={}& \left[ [{\cal G}_{R}]^{-1}-\Sigma(\omega) \right]^{-1},  \\
\Sigma(\omega) ={}& \Sigma_{LR}^\dag \bar{{\cal G}_s}(\omega)\Sigma_{LR} \, ,
\end{align}
\end{subequations}
with ${\cal G}_{R} = {\cal G}_{11}$  and $\Sigma_{LR} = \lambda_s \sigma_z$. 
Thus, the bGF of the voltage biased SC lead enters as
\be 
\bar{{\cal G}_s}(\omega) =  \diag \left( \frac{-\omega_-}{t_s\sqrt{\Delta_s^2 - \omega_-^2}}, \frac{-\omega_+}{t_s\sqrt{\Delta_s^2 - \omega_+^2}} \right),
\ee
where $\omega_\pm = \omega \pm eV$. The Keldish GF appearing in the current written in terms of advanced GFs, using that $G^A_{j j^\prime} = \left[ G^R_{j^\prime j} \right]^\dag$, takes the form
\be 
G^K_{RR}(\omega) = G^\dagger_{RR} \left \{ \Sigma^K + [{\cal G}^\dagger_R]^{-1} {\cal G}^K_R \ [{\cal G}_R]^{-1} \right \} G_{RR} \, ,
\ee
where
\begin{subequations}
\begin{align}
{\cal G}^K_R(\omega) ={}& (1-2f(\omega)) \big({\cal G}^\dagger_R-{\cal G}_R \big) \, ,   \\
\Sigma^K(\omega) ={}& \Sigma_{LR}^\dagger F_\omega \big(\bar{{\cal G}_s}^\dagger-\bar{{\cal G}_s} \big)\Sigma_{LR} \, ,
\end{align}
\end{subequations}
and $F_\omega =\mathbb{I} - 2\diag(f(\omega_-), f(\omega_+))$, being $f(\omega)$ the Fermi-Dirac distribution. Finally, 
\be
\tilde{\Sigma}^{R/A}(\omega) = \Sigma^{R/A}_{ee}(\omega) \Pi_{e} \, , \quad
\tilde{\Sigma}^{K}(\omega) = \Sigma^{K}_{ee}(\omega) \Pi_{e} \, ,
\ee
where we have defined the projector $\Pi_{e} = (\sigma_0+\sigma_z)/2$, as only up-spin electrons from the BCS lead are tunnel coupled to the spinless fermions from the minimal Kitaev chain. Finally, we extract the differential conductance from the current using that $G_{S} = \partial I_{S}/\partial V$ in the $T\rightarrow 0$ limit.

\begin{figure}[t!]
\centering
\includegraphics[width=\columnwidth]{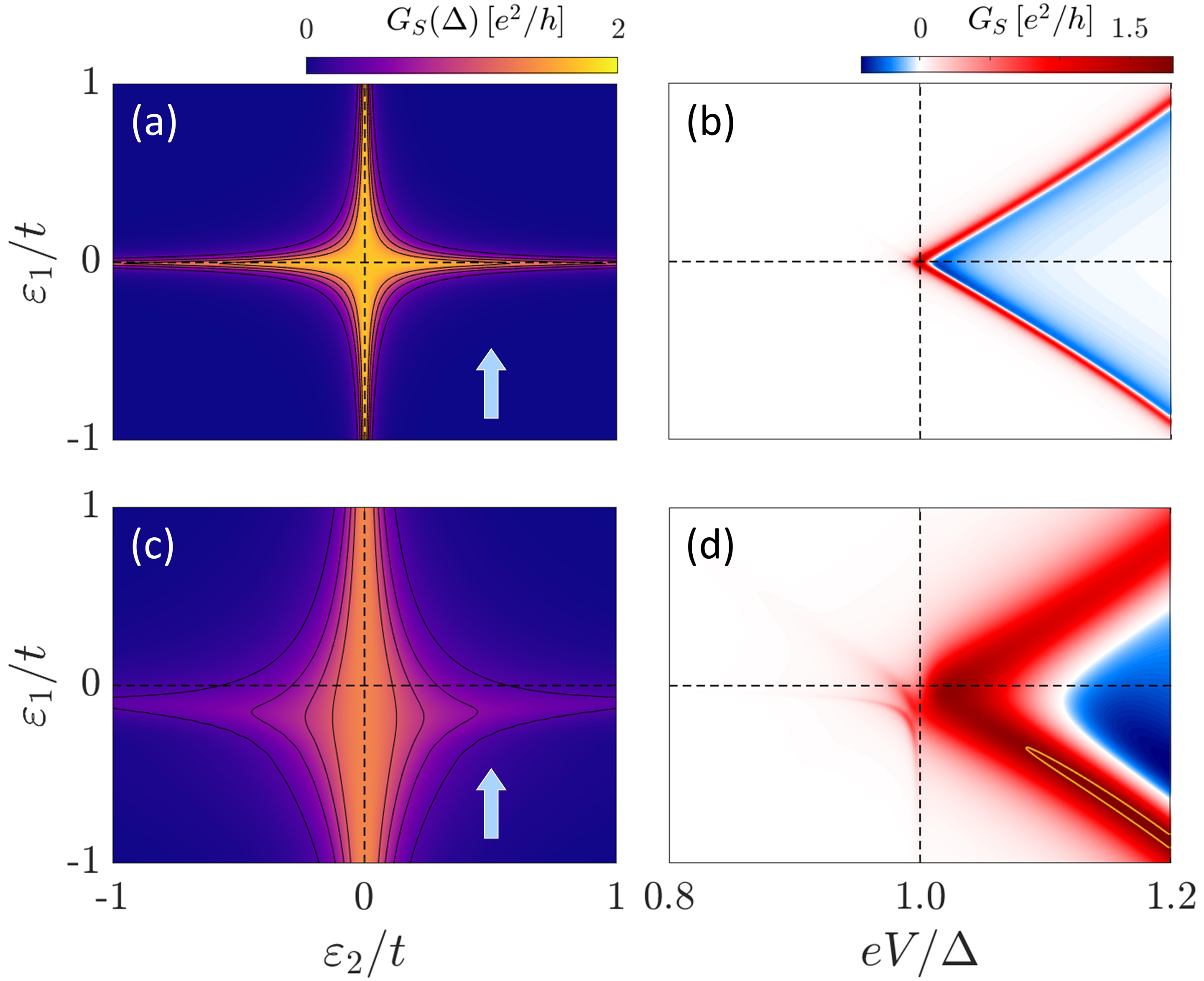}
\caption{Quantum dots' physics characterization in the SC conductance. Left column represents the conductance at fixed voltage bias $eV=\Delta$, where the y-axis (x-axis) represents the local (non-local) QD level for $\lambda_s/t = 0.1$ (a)  and $\lambda_s/t = 0.5$ (c). Right column represents the evolution of the conductance with the level of the first dot for $\lambda_s/t = 0.1$ (b), $\lambda_s/t = 0.5$ (d) and $\varepsilon_2/t = 0.5$ in both cases (marked with blue arrows). The coupling to the normal lead is fixed to $\lambda_n/t = 0.1$. Yellow contour marks the universal height value $G_s/G_0 = (4-\pi)$.}
\label{figS4}
\end{figure}

The small but finite temperature in the calculation allows for thermally excited quasiparticles from the continuum, inducing a broadening term ($\Gamma_{sp}$) of the Majorana zero pole and enabling single particle current by means of inelastic cotunneling. Furthermore, there is an extra relaxation contribution coming from the non-local coupling to the normal lead. It should be noticed that, for a zero energy pole, both resonant Andreev ({\it i.e.}, virtual occupation of the state) and single particle processes ({\it i.e.}, occupation of the state requiring relaxation to empty the state) share the same threshold signal at $e|V|=\Delta$. Notice that, along with the inelastic cotunneling, the hybridization with the SC lead induces a voltage bias dependent broadening ($\Gamma_{e/h}$) of the Majorana zero mode, which may modify the position of the conductance peak related to the threshold. The dominant contribution to the broadening of the subgap state defines two different regimes~\cite{Peng2015, Ruby2015}:

Weak coupling: relaxation faster than tunneling, the transport is obtained as a convolution of the BCS density of states and the zero pole $\Gamma_{sp} \gg \Gamma_{e/h}$. As transport takes place through a discrete level, it is expected to observe conductance peaks going from positive to negative values. 

Strong coupling: tunneling dominates and the BCS density of states modifies the width of the zero pole $\Gamma_{e/h} \gg \Gamma_{sp}$, inducing a bias dependence broadening. Moreover, the negative differential conductance vanishes. 

\begin{figure}[b!]
\centering
\includegraphics[width=\columnwidth]{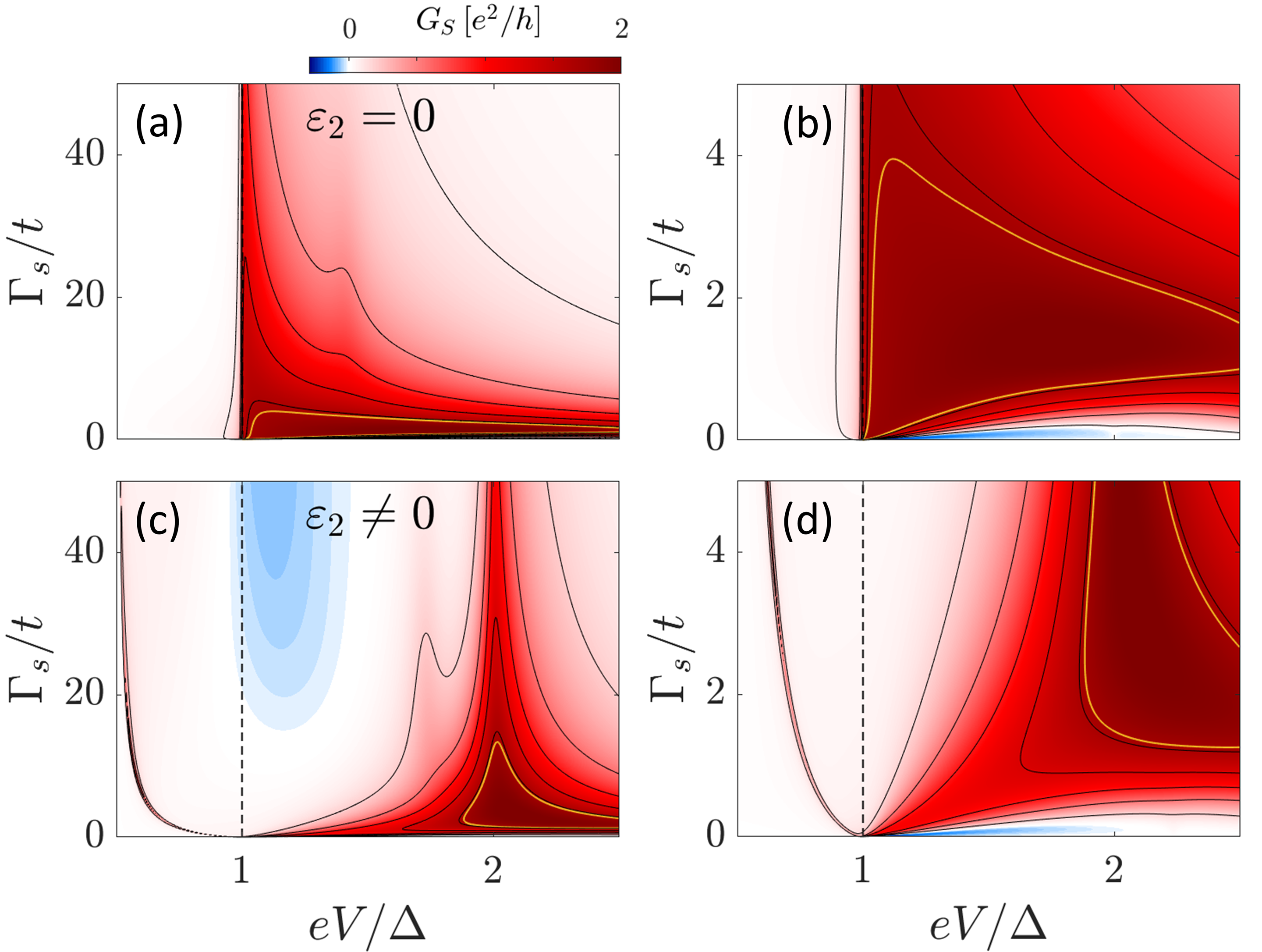}
\caption{Non-local effects in SC transport as a function of the coupling to the SC lead. Upper (Lower) panels, conductance for $\varepsilon_2/t=0$ ($\varepsilon_2/t=1$). Right column represents a close-up image signalling the transition to the ultra-strong coupling. The coupling to the normal lead is set to $\lambda_n/t = 0.1$ in all panels. Yellow contour marks the universal height value $G_s/G_0 = (4-\pi)$.}
\label{figS5}
\end{figure}

In the main text it is mentioned that when detuning the non-local QD it appears a secondary peak dispersing linearly with $\lambda_s$. In this regime when integrating the full range of allowed frequencies $\omega \in [-eV, \, eV]$ to compute the SC current, it is observed the secondary subgap peak not mentioned in previous literature, as they pursued a low energy description near the threshold $e|V| = \Delta_s+\epsilon_0$. This strong coupling regime is a rather ``exotic'' limit in the context of STM experiments on YSR states, however it takes fundamental importance in our platform to explain non-local effects when detuning $\varepsilon_2$, see Figs.~\ref{figS4} \& \ref{figS5}.

\section{Level Physics and Ultra-Strong Coupling} \label{App_E}

Figure~\ref{figS4} illustrates the asymmetrization effect due to the coupling to the SC lead in the QDs' chemical potential physics, thus complementing the analysis made for normal transport in the main text. Fig.~\ref{figS4}(a,c) shows an increasing asymmetry in the conductance peak at voltage bias $e|V| = \Delta$ inherited in the physics of the QD coupled to the SC lead ($\varepsilon_1$). As the asymmetry is increased with the coupling, one loses robustness of the conductance peak when detuning the non-local coordinate ($\varepsilon_2$), Fig.~\ref{figS4}(c). That is, the region showing the universal height value $G_s/G_0=(4-\pi)$ is reduced, and becomes asymmetric in $\varepsilon_1$. Fig.~\ref{figS4}(d) shows non-local subgap structures due to the strong coupling with the SC lead when $\varepsilon_2 \neq 0$. Notice that in the regions where the subgap structure dominates $\varepsilon_1 \gtrsim 0$, the height of the peak above the threshold $eV/\Delta>1$ is further reduced.

Figure~\ref{figS5} illustrates the non-local effects for the SC transport in the ultra-strong coupling regime. First, in the sweet spot and for large couplings $\Gamma_s/t\gtrsim4$ the universal conductance value is never reached again, signalling the transition to the ultra-strong coupling regime, see Fig.~\ref{figS5}(a, b). Moreover, the secondary structure around $eV/\Delta \sim 1.5$ seen at large couplings could be associated to the YSR state when well detached from the continuum. 

Second, when detuning $\varepsilon_2$, it is observed a transition from a subgap dominating regime, where single-particle transport dominates, to another regime around $\Gamma_s/t\gtrsim20$ which reproduces transport phenomenology independent of the coupling to the SC lead, see Fig.~\ref{figS5}(c). As when detuning $\varepsilon_2$, the excited state in the ultra-strong coupling regime becomes pinned to $z_{s,e} \rightarrow \varepsilon_2$, a strong transport signal at the threshold $eV/\Delta \sim 2$ is observed, showing a dispersive satellite peak too. However, Fig.~\ref{figS5}(d) shows that the universal height value is not obtained at the threshold as expected for a true MZM.

\section{Normal Transport}\label{App_F}

For the normal-superconducting case, it is convenient to define a different Gauge with time-independent tunnel couplings $\Sigma_{LR}$ and accounting for the chemical potential difference in the Keldysh GFs. As the coupling only involves one spin species, thus the current can be described in the $2\times2$ Nambu basis
\be
\begin{split}
I_{NS} = \frac{e}{2h} \int \frac{1}{2}\textrm{tr}_N \big \{ \sigma_z \Sigma_{LR} [G_{RR}^{-+} \Sigma_{RL} {\cal G}_L^{+-}  \\
- \, G_{RR}^{+-} \Sigma_{RL} {\cal G}_L^{-+}]   \big \}d\omega \, .
\end{split}
\ee 
We define ${\cal G}_{R} = {\cal G}_{22}$ and ${\cal G}_{L} = {\cal G}_n$, where $\mu_R = \mu_{s} = 0$ and,
\be
\check{G} = 
\bmat G_{LL} && G_{LR} \\ G_{RL} && G_{RR}
\emat = \bmat {\cal G}_L^{-1} && -\Sigma_{LR} \\
-\Sigma_{RL} && {\cal G}_R^{-1} \emat^{-1}.
\ee 
From now on, we will use advanced GF and omit the explicit super-index. The non-equilibrium correlation functions can be expanded using the Langreth rules \cite{Langreth1972, Langreth1976, Cuevas1999} as

\begin{widetext}

\bea
&G^{+-/-+}_{RR} = (\mathbb{I}+ G_{LR}^\dag \Sigma_{LR}){\cal G}^{+-/-+}_R(\mathbb{I}+ \Sigma_{LR}^\dag G_{LR}) + G^\dag_{RR} \Sigma_{LR}^\dag {\cal G}^{+-/-+}_L \Sigma_{LR} G_{RR} \, ,& \nonumber \\
&{\cal G}^{+-}_j = F_j\big({\cal G}_j - {\cal G}_j^\dag \big)\ , \quad {\cal G}^{+-}_j = (F_j-\mathbb{I})\big({\cal G}_j - {\cal G}_j^\dag \big) \, ,&
\eea
where $F_j = f(\omega - \sigma_z\mu_j)\sigma_0$.
Thus, using that $F_R$ is diagonal and commutes with the rest of operators, and that $F_L$ is block diagonal and commutes with operators with no anomalous terms, we obtain
\bea
\kappa_{LR} &=& \sigma_z \Sigma_{LR}(\mathbb{I}+ G_{LR}^\dag \Sigma_{LR})({\cal G}_R - {\cal G}_R^\dag)(\mathbb{I}+ \Sigma_{LR}^\dag G_{LR})\Sigma_{LR}^\dag ({\cal G}_L - {\cal G}_L^\dag)(F_R-F_L) \nonumber \\
& &+ \sigma_z \Sigma_{LR}G^\dag_{RR} \Sigma_{LR}^\dag ({\cal G}_L - {\cal G}_L^\dag) \Sigma_{LR} F_L G_{RR} \Sigma_{LR}^\dag ({\cal G}_L - {\cal G}_L^\dag) \nonumber \\
& &- \sigma_z \Sigma_{LR}G^\dag_{RR} \Sigma_{LR}^\dag ({\cal G}_L - {\cal G}_L^\dag) \Sigma_{LR} G_{RR} F_L \Sigma_{LR}^\dag ({\cal G}_L - {\cal G}_L^\dag) \, .
\eea

To disaggregate the different processes involved in transport, we explicitly expand in Nambu space the different terms of $\kappa_{LR} = \kappa_{LR}^I+\kappa_{LR}^{II}+\kappa_{LR}^{III}$ to compute the trace, where we have assumed that $\Sigma_{LR} = t_e \sigma_z$ and $t_e = \lambda_n$. Then, for example
\be
\kappa_{LR}^I = \sigma_z \Sigma_{LR} (\mathbb{I}+ G_{LR}^\dag \Sigma_{LR})A(\mathbb{I}+ \Sigma_{LR}^\dag G_{LR})\Sigma_{LR}^\dag B (F_R-F_L) \, , 
\ee
thus,
\bea
\kappa_{LR,ee}^I &=& t_e [ (\mathbb{I}+ G^\dag_{LR,ee}t_e)A_{ee} - G^\dag_{LR,he}t_e A_{he} ] (\mathbb{I}+ t_e^\dag G_{LR,ee})t_e^\dag B_{ee} [n(\omega)-n(\omega-eV)] \nonumber \\
&& -t_e [ (\mathbb{I}+ G^\dag_{LR,ee}t_e)A_{eh} - G^\dag_{LR,he}t_e A_{hh} ] t_e^\dag G_{LR,he}t_e^\dag B_{ee} [n(\omega)-n(\omega-eV)] \, ,
\eea
where $A=({\cal G}_R - {\cal G}_R^\dag)$, and $B=({\cal G}_L - {\cal G}_L^\dag)$. From this term it is possible to extract the explicit expression for the different quasiparticle contributions to the electron current, such
\begin{subequations}
\begin{align}
\kappa_{Q_{1,ee}} ={}& t_e (\mathbb{I}+ G^\dag_{LR,ee}t_e)A_{ee} (\mathbb{I}+ t_e^\dag G_{LR,ee})t_e^\dag B_{ee} [n(\omega)-n(\omega-eV)] \nonumber \\
& + t_e G^\dag_{LR,he}t_e A_{hh} t_e^\dag G_{LR,he}t_e^\dag B_{ee} [n(\omega)-n(\omega-eV)] \, , \\
\kappa_{Q_{2,ee}} ={}& - t_e G^\dag_{LR,he}t_e A_{he}
 (\mathbb{I}+ t_e^\dag G_{LR,ee})t_e^\dag B_{ee} [n(\omega)-n(\omega-eV)] \nonumber \\
& - t_e (\mathbb{I}+ G^\dag_{LR,ee}t_e)A_{eh} t_e^\dag G_{LR,he}t_e^\dag B_{ee} [n(\omega)-n(\omega-eV)] \, . \\
\kappa_{A,ee} ={}& t_e G^\dag_{RR,he} t_e^\dag B_{hh} t_e G_{RR,he} t_e^\dag B_{ee}[n(\omega+eV)-n(\omega-eV)] \, .
\end{align}
\end{subequations}
where the Andreev contribution to current is obtained from the remaining terms $\kappa_{A,ee} = \kappa_{LR,ee}^{II} + \kappa_{LR,ee}^{III}$. Equivalently, the hole contributions to the current can be computed as
\begin{subequations}
\begin{align} 
\kappa_{Q_{1,hh}} ={}& -t_e (\mathbb{I}- G^\dag_{LR,hh}t_e)A_{hh} (\mathbb{I}- t_e^\dag G_{LR,hh})t_e^\dag B_{hh} [n(\omega)-n(\omega-eV)] \nonumber \\
& - t_e G^\dag_{LR,eh}t_e A_{ee} t_e^\dag G_{LR,eh}t_e^\dag B_{hh} [n(\omega)-n(\omega-eV)] \, , \\
\kappa_{Q_{2,hh}} ={}& - t_e G^\dag_{LR,eh}t_e A_{eh}
 (\mathbb{I}- t_e^\dag G_{LR,hh})t_e^\dag B_{hh} [n(\omega)-n(\omega-eV)] \nonumber \\
& - t_e (\mathbb{I}- G^\dag_{LR,hh}t_e)A_{he} t_e^\dag G_{LR,eh}t_e^\dag B_{hh} [n(\omega)-n(\omega-eV)] \, ,  \\
\kappa_{A,hh} ={}& t_e G^\dag_{RR,eh} t_e^\dag B_{ee} t_e G_{RR,eh} t_e^\dag B_{hh}[n(\omega+eV)-n(\omega-eV)] \, .
\end{align}
\end{subequations}

In the zero temperature limit, as in the linear response approach for an Ohmic contact the only explicit dependence in the bias voltage comes from the Fermi distribution, therefore 
\be 
f^\prime_L = \frac{\partial}{\partial V} f(\omega \mp eV) \rightarrow \frac{\partial}{\partial V} (1- \Theta(\omega \mp eV)) = \pm e\delta(\omega \mp eV) \, ,
\ee 
using that the partial derivative of the Fermi distribution in the zero temperature limit is a Dirac's delta. Finally, the differential conductance at zero temperature takes the form
\bea
G_{NS} = -\frac{e^2}{2h} && \textrm{tr} \big \{ \tilde{\kappa}_{A,ee}(-eV) +\tilde{\kappa}_{A,hh}(-eV) - \tilde{\kappa}_{Q_1,hh}(-eV)-\tilde{\kappa}_{Q_2,hh}(-eV) \nonumber \\ &&+\tilde{\kappa}_{A,ee}(eV) +\tilde{\kappa}_{A,hh}(eV) + \tilde{\kappa}_{Q_1,ee}(eV)+\tilde{\kappa}_{Q_2,ee}(eV) \big \} \, ,
\eea
where $\tilde{\kappa}_\mu$ represents the contributions to the current without the Fermi-Dirac terms, and the trace may take into account other possible orbital degrees of freedom. 

\end{widetext}

\section{Ultra-Strong Coupling Regime }\label{App_G}

\begin{figure}[t!]
\centering
\includegraphics[width=\columnwidth]{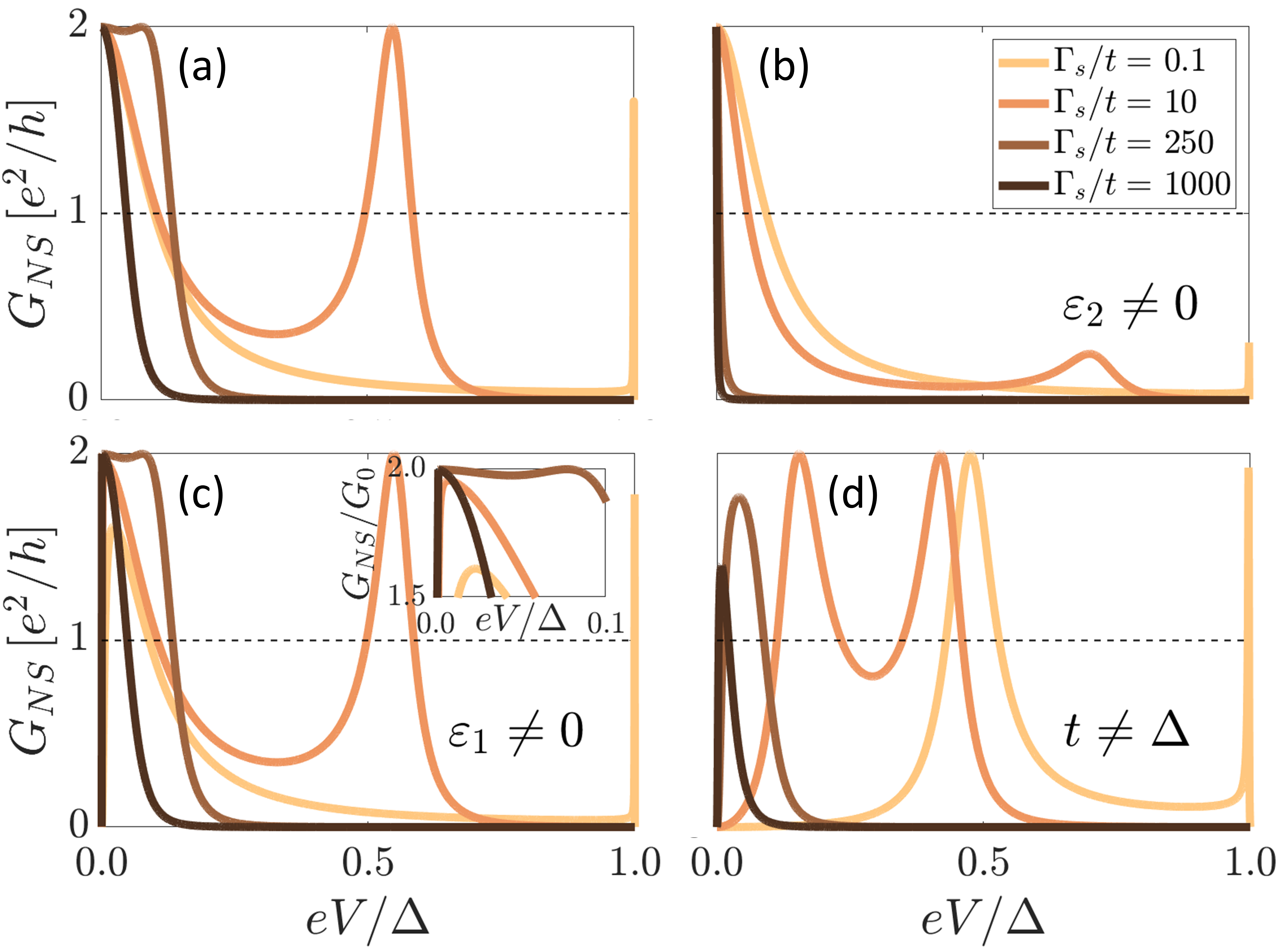}
\caption{
Interplay between YSRs and MZMs in normal conductance as a function of the coupling to the SC lead. (a) Conductance cuts in the sweet spot configuration $t/\Delta =1$ and $\varepsilon_1 = \varepsilon_2 = 0$. The rest of the panels shows situations out of the sweet spot such (b) $\varepsilon_2 \neq 0$ (c) $\varepsilon_1 \neq 0$ and  (d) $t/\Delta = 0.5$. Inset shows a close-up image around zero-bias. The coupling to the normal lead is set to $\lambda_n/t = 1$ in all panels. }
\label{figS6}
\end{figure}

Figure~\ref{figS6} illustrates various normal conductance profiles for different values of $\Gamma_s$ considering the interplay between MZM and YSR states when the system deviates from the sweet spot. In all cases we consistently see that all subgap states tend towards zero energy for large enough couplings to the SC lead. Therefore, the gap to excited states tends towards zero with the coupling to the superconducting lead.

As discussed in the main text, the normal conductance at the sweet spot in Fig.~\ref{figS6}(a) shows a quantized YSR states coexisting with a quantized zero-bias peak. Remarkably, in the ultra-strong coupling regime the quantized zero-bias conductance peak persist, despite the overlap of YSR and Majorana states.

Figure~\ref{figS6}(b,c) shows the YSR-MZM physics when one of the QDs in the chain is detuned. In the case of $\varepsilon_2 \neq 0$, shown in FIG.~\ref{figS6}(b), it should be noticed the absence of quantized transport response for the YSR states, resulting in a vanishing normal transport signal for larger couplings. Nonetheless, the quantized zero-bias peak remains intact, indicating the robustness of the MZM peak against this detuning. Conversely, FIG.~\ref{figS6}(c), considers the case where $\varepsilon_1 \neq 0$. Here, a distinct quantized peak associated with the YSR states emerges, accompanied by a zero-bias dip. As the coupling to the SC lead is further increased, the low-bias conductance exhibits an increase as the YSR energy decreases. When both subgap states merge, the quantized values area reached for both broadened peaks, while the zero-bias dip is preserved.

Lastly, Fig.~\ref{figS6}(d) addresses the $t\neq\Delta$ case, revealing the MZM as a non-zero energy subgap state that displays quantized conductance. In the strong coupling, the YSR state begins to detach from the continuum. This non-trivial subgap structure aligns with the evolution of poles detailed in App.~\ref{App_A}. When the subgap states are well isolated between them and the continuum, they show quantized conductance behaviour. However, at even higher values of $\Gamma_s$, these states merge as they converge toward zero-bias. This merging effect results in a pronounced zero-bias dip and a decrease in the overall transport signal as the coupling to the SC lead continues to increase, illustrating the complex interplay between both subgap states.

\begin{thebibliography}{79}
\bibitem [{foo()}]{footnote1}%
  \BibitemOpen
  \href@noop {} {}\bibinfo {note} {It should be noticed that in a realistic system there are present both residual p-wave correlations and residual s-wave. However for a fully spin-polarized minimal Kitaev chain it is an exact approximation to neglect the anomalous part of the BCS probe Green's function. Therefore, the conservation of current requires virtual transitions to the SC lead.}\BibitemShut {Stop}%
\end{thebibliography}

\providecommand{\noopsort}[1]{}\providecommand{\singleletter}[1]{#1}%
%


\end{document}